\documentclass[a4paper,11pt]{article}
\pdfoutput=1 

\usepackage{jheppub} 

\usepackage[T1]{fontenc} 

\usepackage{tikz}
\usepackage{booktabs}
\usepackage{cellspace}
\usepackage{nicefrac}
\usepackage{url}
\usepackage{hyperref}

\newcommand{\E}{\mathrm{e}}
\newcommand{\Exp}{\mathrm{exp}}

  {\left\lbrace\begin{array}{@{}l@{}}}%
  {\end{array}\right.}

\title{\boldmath RG and logarithmic CFT multicritical properties \\ of randomly diluted Ising models}

\author[a]{R.\ Ben Al\`i Zinati}
\author[b]{and O.\ Zanusso,}

\affiliation[a]{Sorbonne Universit\'e \& CNRS, Laboratoire de Physique Th\'eorique de la Mati\`ere Condens\'ee,\\ F-75005,  Paris, France}
\affiliation[b]{Universit\`a di Pisa and INFN - Sezione di Pisa, Largo Bruno Pontecorvo 3, I-56127 Pisa, Italy}

\emailAdd{riccardo.ben\_ali\_zinati@sorbonne-universite.fr}
\emailAdd{omar.zanusso@unipi.it}

\abstract{
We discuss how a spin system, which is subject to quenched disorder, might exhibit
multicritical behaviors at criticality if the distribution of the impurities is arbitrary.
In order to provide realistic candidates for such multicritical behaviors,
we discuss several generalizations of the standard randomly diluted Ising's universality class
adopting the $\epsilon$-expansion close to several upper critical dimensions.
In the presentation, we spend a special effort in bridging between CFT and RG results
and discuss in detail the computation of quantities, which
are of prominent interest in the case of logarithmic CFT.
}

\begin{document}
\maketitle

\section{Introduction}
\label{sec:intro}

It is a well-known fact that statistical mechanics systems such as lattice spin models may exhibit critical behavior by undergoing a second order phase transition
if some external parameters such as the temperature are opportunely tuned. This is the case, for example,
of the lattice Ising model, which has diverging correlation length at a special critical temperature $T_c$
and which falls into the same universality class of a single-component $\phi^4$ field theory.
It is less-known, however, that many of the same systems still exhibit critical behavior even in the presence of some
probability-driven dilution of the lattice sites.
A physical situation in which dilution is important happens when there are impurities in the system,
for example non-magnetic sites in a ferromagnetic lattice, and these impurities are distributed on the lattice according
to some probability.

One could correctly argue that, if not many impurities are present, the diluted system
must behave similarly to the pure (non-diluted) one.
The truth is actually more surprising:
for sufficiently low concentration and, to some extent,
independently on the distribution of impurities, the system undergoes a \emph{different}
second order phase transition with lower $T_c$ if the thermodynamical exponent $\alpha$ is positive \cite{Harris:1974}.
Under the assumption that system's impurities change according to a timescale
that is much longer than the time to reach equilibrium, the randomness
introduced as a result must be taken into account through quenched averages
over the disorder.
This practically means that disorder's averages are performed at the very end
of any correlator's computation and
the difference between the pure and diluted systems is quantified by the final quenching.

Much of previous work on the effect of disorder has concentrated attention
to dilution of the spins of the lattice Ising model assuming a Gaussian uncorrelated distribution
of the impurities.
In this case, the difference between the critical exponents of pure and diluted systems
has been firmly established by field-theoretical and lattice methods (see refs. in \cite{Pelissetto:2000ek}).
Less is known on what happens if other spin systems, such as the tricritical Ising model,
are faced with random dilution, or, alternatively,
if the distribution of impurities is changed to one that depends on more parameters.
One might wonder if both the above changes might lead to \emph{multicritical} behavior
of the diluted system, essentially by introducing new parameters to be tuned to criticality.
A simple answer to this question can be achieved by classifying and discussing
critical field-theoretical models representing diluted systems by means of renormalization group (RG) methods.

On the practical side, the effect of dilutions can be taken into account by introducing $N$ non-interacting
replicas of the system and integrating over the distribution of impurities
(this \emph{subterfuge} is the so-called replica method) \cite{cardy1996scaling}.
The net effect of the procedure is that, after integration, the copies of the system become interacting
and, for a randomness that modifies the energy-per-site operator, the replicated system has symmetry enhanced
by the discrete group $H_N$ (the symmetries of the hypercube), if compared to the symmetry of the pure Ising system.
An interesting aspect of the procedure is that quenched averages can be computed
exactly like traditional path-integral averages if the limit $N\to 0$ is taken.
Therefore, the last step in quenching requires an analytic continuation of the results for arbitrary $N$.

It is worth sidestepping for a moment to discuss the interpretation of the analytic continuation in $N$
from a conformal field theory (CFT) point of view. First recall that, rather generally, critical points of statistical systems
can be associated to conformal field theories. This implies that the simple space- and scale-invariance
of the critical system (invariance under the action of the dilatation operator)
is actually enhanced by the full group of conformal transformations. At the critical point CFT methods can be used
to determine various quantities of interest, including the scaling dimensions of operators, which
are in correspondence with critical exponents of the system.
This is certainly true for arbitrary natural-valued $N$, in which there are, for example, $N$ total CFT primary operators
with maximum scaling dimension besides the identity operator related by the symmetry $H_N$.
Less trivial to visualize the limit $N\to 0$, because this would actually correspond to no fields at all.

Amazingly, the limit can be formalized with surprising results. First of all, it is important to stress that
the operator content can be classified for arbitrary $N$ adopting the representation theory of the group $H_N$.
This allows one to work with general $N$ at any step of the computation, without limitations from considering
specific integer values. Then, upon continuation in $N$, some operators of the spectrum, which are generally distinguished as realizing irreducible representations of $H_N$, have degenerate scaling dimensions for
$N\to 0$. This happens because, in the limit, the dilatation operator does not commute anymore with
the action of the group $H_N$, and has the effect to produce a so-called logarithmic CFT (log-CFT).
In practice, the energy operator and a tensor operator of rank two have colliding scaling dimensions for $N\to 0$,
implying that their correlators are not diagonal anymore, and the limit produces a specific structure
which ultimately results in a logarithmic correction to the traditional power law structure of the CFT correlators \cite{Gurarie:1993}.
An outstanding result is that these logarithmic contributions introduce new universal coefficients
and can be related to meaningful observables in terms of product and ratios of quenched averages \cite{Cardy:1999, Cardy:2013}.

In the light of the existence of new universal quantities,
the computation of the logarithmic effects in the CFT spectrum gives a brand new
arena in which it is interesting to compare RG and CFT based results.
The challenge is to understand from an RG point of view all the intricacies of the log-CFT
approach, but also to prove once more the validity of the RG approach by showing
that it can easily accommodate the necessary analytic continuation in $N$.
This last step is of course greatly helped by the use of representation theory of the group $H_N$.
Using RG, quantities relevant for the log-CFT framework could be initially computed by means of some
perturbative expansion such as the $\epsilon$-expansion, and possibly improved in the future
by non-perturbative methods.

In this paper, we discuss several multicritical generalizations of the diluted spin system universality class,
understood as the $N\to 0$ limit of a replicated field theory. We do this by investigating critical Ginzburg-Landau-like
Hamiltonians with $H_N$ symmetry and by constraining them to be coming from a replicated system.
The critical models have associated upper critical dimensions, which can be deduced on the basis of dimensional analysis, so we study the first interesting possibilities, before the size of the computations gets out of hand.
All the nitty-gritty details of the computations, that are necessary to obtain the results presented in this paper,
will be presented in a future paper, in which we will also address the general case with arbitrary $N$ \cite{bcz-to-appear}.
%

The rest of the paper is structured as follows:
we pedagogically review the basic steps of using replicas to treat disorder, make the case for the multicritical generalizations by enhancing the possible distributions for the disorder in Sect.~\ref{sec:hypercubicandreplica},
we briefly discuss the connection with the formalism of log-CFT highlighting the simplest logarithmic observable
in Sect.~\ref{sec:logarithms},
and we provide some further notion on criticality and renormalization group methods in Sect.~\ref{sec:method}.
The explicit results are presented in the subsequent sections:
In Sects.~\ref{sec:dc3}~to~\ref{sec:dc10/3} we discuss three multicritical generalizations
of the randomly diluted Ising model.
For each model, we pay special attention to the computation of the relevant scaling dimensions
in the language of log-CFT.
In App.~\ref{app:betas}, we list the most important RG quantities, which are needed to obtain our results.
In App.~\ref{sec:dc4}, we review the $\phi^4$-like model in $d=4-\epsilon$ in order to make
our discussion complete (in passing, we collect several quantities which are otherwise scattered in the literature).

\section{Hypercubic and replica}
\label{sec:hypercubicandreplica}
The influence of quenched frozen-in structural disorder on the critical behavior of
a magnetic model
can be accounted for by diluting the system with non-magnetic impurities randomly distributed over the original lattice.
For a first concrete example, consider an Ising ferromagnet in which a fraction of the spins is replaced by vacancies; in this form of site-dilution, the disorder is explicitly present in the Hamiltonian under the form of random variables $m_i$ taking the values $0$ or $1$ depending on whether or not the site $i$ is vacant
\begin{equation}\label{eq:microHamiltonian}
  \mathcal{H}[\{m_i\}] = - J \sum_{\langle i, j \rangle} m_i m_j s_i s_j \,,
\end{equation}
and the spins $s_i=\pm 1$ are the degrees of freedom. For a second example, consider a random bond dilution in which randomness is moved from the lattice sites to the coupling constants resulting in the Hamiltonian
$\mathcal{H}[\{J_{ij}\}] = -\sum_{\langle i j \rangle} J_{ij}s_is_j$. The disorder is completely specified by a probability distribution function, $P[\{m_i\}]$ in the case of site-dilution
and $P[\{J_{ij}\}]$ in the case of bond
randomness.\footnote{%
For completeness, we mention that bond and site randomness are not the only types of disorder that can be encountered experimentally.
Random fields that couple linearly to the magnetic moments can also be considered.
These are described by the so-called random field Ising model (RFIM),
which displays a form of disorder that is essentially different from the aforementioned ones,
mostly because the external random field can break the $\mathbb{Z}_2$ symmetry
w.r.t.\ spin flip.
RFIM is a very interesting and not yet fully understood model,
and we point the interested reader to the excellent reviews \cite{Nattermann, Dotsenko3,Natterman2}
}
In either case,
it is natural to expect that the presence of impurities
manifests itself by lowering the critical temperature at which the phase transition to an ordered phase occurs.
Sufficiently close to the critical point of the pure theory, at which an effective continuum description is possible,\footnote{%
For example by means of a Hubbard-Stratonovich transformation that allows to sum over the spins.}
quenched disorder can be described in terms of small random spatial fluctuations of the effective temperature.
Namely, we define the action of the diluted model as
\begin{equation}
 S[\phi] = S_0[\phi] + \int {\rm d}^dx \, m(x) E_1(x) \,,
\end{equation}
in which  $S_0[\phi]$ is the action of the pure magnetic model,
and $m(x)$ is a random variable coupled to the energy density operator $E_1=\phi^2$.
The energy operator has an additional label for reasons that are clarified in a moment.

Intuitively, physical quantities like the free energy or correlation functions should not depend on a specific realization of the disorder, i.e.\  they should be \emph{self-averaging} over the realizations of the disorder according to its distribution function. Phrased differently, self-averaging quantities are such that, in the thermodynamic limit, they are equal to their quenched expectation value \cite{cardy1996scaling, mezard1987spin}.

The quenched average is greatly simplified by replicating the field into $N$ non-interacting copies $\phi \to \phi_a=\{\phi_1,\phi_2,\dots,\phi_N\}$, and then integrating over the disorder.
The rationale behind this procedure stems from the following identity
\begin{equation}
  \log Z = \lim_{N \to 0} \frac{Z^{N}-1}{N}\,,
\end{equation}
which recasts the difficult problem of averaging the logarithm of the partition function into the easier one of averaging the partition function of $N$ independent copies or \emph{replicas} of the original system.
Evidently, analytic continuation to $N\to 0$ is required for full exploitation of the method.\footnote{%
An alternative to this approach would be the strategy in which the same limit is reached by having a finite number
of bosonic and fermionic degrees of freedom which cancel out. This leads to structures with manifest supersymmetry \cite{Cardy:1985}.}
Notice that, likewise the field, local pure operators are replicated too; for example the energy density operators for each replica are $E_a \equiv (\phi_a)^2$.

For the purpose of integrating over disorder,
let us consider a generic random variable $X$ with probability distribution function (p.d.f.) $P(X)$;
the average of a function $f(X)$ is
expressed as $\overline{f} \equiv \int {\rm{d}}X P(X) f(X)$.
One can construct the generating function
\begin{equation}\label{eq:cumulantsexp}
    G(\xi) \equiv \overline{\E ^ {\xi X}}
= \Exp \left({\sum_{j=1} ^\infty \frac{\xi^j}{j!} \kappa_j} \right) \,,
\end{equation}
in which $\kappa_j$ are the cumulants of the probability distribution $P(X)$.
We can treat the disorder in the replicated quenched-averaged partition function by performing the same cumulant expansion with the correspondence $X\leftrightarrow m(x)$ and $\xi \leftrightarrow -\sum_a E_a(x)$, namely
\begin{equation}\label{eq:avgZ}
\begin{split}
  \overline{Z^N}
  & = \int ({\rm d}m) P[ m(x) ] \int ({\rm{d}}\phi_a)  \E^{-\sum_{a} \left[ S_{0,a} + \int {\rm{d}}^d x ~ m(x) E_a(x)\right]} \\
  & = \int ({\rm{d}}\phi_a)  \E^{-\sum_{a} \left[ S_{0,a} + \kappa_1 \int {\rm{d}}^d x ~ E_a(x)\right]
    + \frac{1}{2!} \kappa_2 \int {\rm{d}}^d x ~ \sum_{a , b} E_a(x) E_b(x) + \dots}\,,
\end{split}
\end{equation}
in which we only assumed that impurities are just locally correlated.
The process of averaging over the quenched disorder results in the different replicas becoming \emph{interacting}
and the strength and form of their interactions depend on the details of the probability distribution.

For later purpose, it is interesting to show the truncated form of the replicated action $S_R$,
which retains all the terms up to order $\phi^6$. This is given by
\begin{align}\label{eq:replicated-action}
  S_R[\phi] = &  \sum_a S_{0,a} + \int {\rm{d}}^d x \Bigl\{ \kappa_1 \sum_a E_a(x)
  - \frac{\kappa_2}{2!} \sum_{a,b} E_a(x)E_b(x) + \frac{\kappa_3}{3!} \sum_{a,b,c} E_a(x) E_b(x) E_c(x) \Bigr\} \notag\\
  = & \int {\rm{d}}^d x ~ \Bigl\{
  \sum_a \Bigl[ \frac{1}{2} (\nabla \phi_a)^2 + \mu^2 E_a(x) + g_1 E_a^2(x) + w_1 E_a^3(x)\Bigr]  \notag \\
  &  + g_2 ~ \sum_{a \neq b} E_a(x) E_b(x) + w_2 \sum_{a \neq b} E_a^2(x)E_b(x) + w_3 \sum_{a\neq b \neq c} E_a(x) E_b(x) E_c(x) \Bigr\} \,,
\end{align}
in which in the last step we explicitly separated the decoupled replica parts from the coupled ones,
and we also introduced effective couplings $\{\mu, g_{1,2}, w_{1,2,3}\}$.
The effective couplings are combinations of the pure ones and of the cumulants,
for example $\mu^2$ is the sum of $\kappa_1$ and of the analog term present in $S_{0,a}$, which
we have not shown for brevity.

For concreteness one can consider the traditional example \cite{Grinstein:1976} of a regular lattice in which each site is  occupied by a spin variable with probability $p$ or it is empty with probability $1-p$. This situation corresponds to the following bi-modal distribution function
\begin{equation}\label{eq:bimodalpdf}
  P[m(x)] = \prod_x \left[ p~\delta_{m(x),1} + (1-p)~\delta_{m(x),0} \right]\,.
\end{equation}
It is straightforward, in this case, to compute the first cumulants and show that they are nonzero for almost all values of $p$. In the cases of the microscopic Hamiltonian \eqref{eq:microHamiltonian} and site dilution \eqref{eq:bimodalpdf}, it is possible to construct the effective Hamiltonian corresponding
to the replicated action \eqref{eq:replicated-action}, in which the correspondence
between the microscopic parameters of the lattice model and those of the continuum model is explicit, see Ref.~\cite{Grinstein:1976}.
Notice that, even if for some probability distribution functions some terms might not be generated by the random variable (e.g.\ those related to odd cumulants for a symmetric p.d.f.), the iteration of the renormalization group procedure
will likely produce an action such as \eqref{eq:replicated-action} after the first iteration.

The replicated action is invariant under the group $S_N$ of permutations of the replicas, but each replica shares the original Ising-like $\mathbb{Z}_2$ symmetry, therefore the overall symmetry group characterizing the replicated action $S_R$ is the hypercubic point group $H_N \simeq (\mathbb{Z}_2)^N \rtimes S_N$.
In other words, the system has now the symmetries of the hypercubic model.
Quenched averages can actually be constructed with $S_R[\phi]$ and arbitrary $N$,
but they involve correlators normalized by $N$-powers of the partition functions.
The limit $N\to 0$, however, gives quenched correlators as standardly normalized correlators of $S_R$.
In the following the strategy will be to consider the theory for arbitrary analytically continued $N$,
and perform the limit as the very last step.

It is important to mention here that, contrary to pure systems for which the ground state is unique,
random systems might have many more local minima of the energy.
Therefore, the standard RG approach may fail in taking into account
the several contributing configurations of the replicated action \eqref{eq:replicated-action},
which may also cause the spontaneous breaking of the replica symmetry.
This problem, which is very well-known in the spin-glass community, motivated some authors \cite{Dotsenko1, Dotsenko2} to reconsider the RG study of randomly diluted spin systems
in terms of the Parisi's replica symmetry breaking (RSB) scheme,
which has been specifically developed to deal with systems exhibiting
several local minima configurations. Further investigations, however,
proved that the replica-symmetric FP solution in $d=3$
is stable against the switching on of possible replica symmetry breaking terms \cite{Prudnikov1}.

Even a simple distribution such as \eqref{eq:bimodalpdf} turns on \emph{all} cumulants and therefore,
in principle, all possible interactions of the replicated action \eqref{eq:replicated-action}.
We use this notion to argue that multicritical generalizations of the hypercubic model, as hinted
from our formula for the replicated action, are not only interesting in their own right,
but important for the sake of constructing models of quenched average with arbitrary
interactions and distributions of the impurities.
Not much is known about these multicritical generalizations;
for example the possibility of RSB in the multicritical case has, to our best knowledge,
never been invistigated.
Starting from Sect.~\ref{sec:dc3},
we characterize and discuss the first few interesting examples of critical and multicritical
models with hypercubic symmetry.

\section{Logarithms and logarithmic CFTs}
\label{sec:logarithms}

Now we feel the urge to make a connection with a language more akin to that of CFT.
In the previous section, we established the possibility to work with arbitrary $N$ before continuing to the limit $N\to 0$,
which is
quite advantageous when it comes to constructing interesting observables, because we can
simplify field's multiplets in terms of irreducible representations of the hypercubic group.
Here we concentrate on observables built with two copies of $\phi$ and no derivatives,
but a much more general discussion is possible \cite{Cardy:2013,Vichi:2016}.
The study of the irreducible representations of $H_N$ gives three distinct operator multiplets at the quadratic level
for arbitrary $N$
\begin{equation}\label{eq:irreps}
 S = \sum_a \phi_a^2 \,, \qquad\qquad  X_{ab} = \phi_a^2- \phi_b^2\,, \qquad\qquad Y_{ab} =\phi_a\phi_b\,,
\end{equation}
in which $S$ is obviously the singlet, $X$ is an antisymmetric two-tensor, and $Y$ is a symmetric one.
The naive expectation is that, for arbitrary values of $N$ the dilatation operator ${\cal D}$ should commute with the
action of the elements of the group $H_N$, implying that a scaling operator ${\cal O}(x)$
can carry a label for ${\cal D}$ (a.k.a.\ the scaling dimension) as well as an irreducible representation of $H_N$.
From the CFT point of view, the corresponding states would be of the form $\left|\Delta,R\right>\simeq{\cal O}\left|0\right>$ with $R$ ranging over the irreducible representations, for example $R=S,X,Y,\dots$ as in \eqref{eq:irreps}.\footnote{%
One can adopt radial quantization to make the connection between states and operators more formal.}

A general property, that we confirm repeatedly in the next sections, is that the scaling dimensions
of the operators $S$ and $X$ degenerate in the limit $N\to 0$. One way to understand this is that
the copies can be thought to become completely indistinguishable in the limit.
To clarify the connection with the previous notation and to make contact with previous literature
(modulo some normalization), we make the first important identification:
the singlet $S$ is the energy operator of the replicated system up to an overall constant,
so we define $E\equiv\frac{1}{N}S=\frac{1}{N}\sum_a E_a$.
The second identification is only slightly less trivial. Given that the tensor $X$ has two indices,
we first take the trace over one index to construct a vector $\tilde{E}_a\equiv \frac{1}{N}\sum_b X_{ab}=E_a-E$.
This vector has the same scaling dimension as $X$ and, by construction, is traceless $\sum_a \tilde{E}_a=0$ \cite{Cardy:1999}.

The two point correlators of the operators $E$ and $\tilde{E}$ can be simplified using replica symmetry,
that is, the symmetry of the subgroup $S_N$ acting on the labels $a=1,\dots,N$. In particular, the correlators can easily be simplified
in terms of those of only two distinguished copies $a=1,2$, which is the minimum number of necessary copies.
The result is
\begin{eqnarray} \label{eq:replica-correlators}
 \langle E(x) E(0)\rangle &=& \frac{1}{N} \langle E_1(x) E_1(0)\rangle + \frac{N-1}{N}\langle E_1(x) E_2(0)\rangle \,,\\
 \langle \tilde{E}(x) \tilde{E}(0)\rangle &=& \frac{N-1}{N} \langle E_1(x) E_1(0)\rangle -\frac{N-1}{N}\langle E_1(x) E_2(0)\rangle \,.
\end{eqnarray}
The interesting aspect of this manipulation is that the correlators on the right hand side
can be used to give the quenched average of both connected and disconnected energy-energy correlation functions,
which are defined as
\begin{eqnarray} \label{eq:quenched-correlators}
 \overline{\langle E(x) E(0)\rangle} =\lim_{N\to 0} \langle E_1(x) E_1(0)\rangle\,,  &\qquad \quad&
 \overline{\langle E(x) \rangle \langle E(0)\rangle} =\lim_{N\to 0}\langle E_1(x) E_2(0)\rangle
\end{eqnarray}
and are observables of the diluted system.
We thus solve for the correlators on the r.h.s.\ of \eqref{eq:replica-correlators} inside \eqref{eq:quenched-correlators},
to obtain in the limit two very important observables:
\begin{equation}
\begin{split}
 &\overline{\langle E(x) E(0)\rangle} -\overline{\langle E(x) \rangle \langle E(0)\rangle}
  = \lim_{N\to 0} N \langle \tilde{E}(x) \tilde{E}(0)\rangle \,,\\
 &\overline{\langle E(x) \rangle \langle E(0)\rangle}
  = \lim_{N\to 0} \Bigl\{\langle E(x) E(0)\rangle +\langle \tilde{E}(x) \tilde{E}(0)\rangle \Bigr\}\,.
\end{split}
\end{equation}
In manipulating these expressions, we assume that correlators can have poles at $N=0$, but are otherwise regular.
In fact, the first line is nonzero only if the correlator of $\tilde{E}$ is singular in the limit,
which has been confirmed in practice by CFT methods.

Now we use the scaling properties of the hypercubic model at criticality for general $N$
to deduce the scaling form of the correlators
\begin{eqnarray}
 \langle E(x) E(0)\rangle \sim   \frac{1}{N} A(N) \left|x\right|^{-2\Delta_E} \,, &\quad &
 \langle \tilde{E}(x) \tilde{E}(0)\rangle \sim   \frac{N-1}{N} \tilde{A}(N) \left|x\right|^{-2\Delta_{\tilde{E}}}\,,
\end{eqnarray}
in which $\Delta_E$ and $\Delta_{\tilde{E}}$ are the $N$-dependent scaling dimensions of the operators $E$
and $\tilde{E}$ respectively.
The structure of the overall coefficients can be inferred from the analysis of the replicated correlators:
the functions $A(N)$ and $\tilde{A}(N)$ are regular in the limit $N\to 0$.
Evidently, the normalization of the correlators is singular in the limit, which can be proven using CFT methods and requiring consistency.
The presence of these poles is actually very important from the point of view of CFT,
because they generate the most interesting aspects of the quenched limit, as we shall see briefly.

Recall now that the two operators become degenerate in the limit $N\to 0$:
consistency of the correlators requires that $A(0)=\tilde{A}(0)$ and $A'(0)=\tilde{A}'(0)$.\footnote{%
Actually, the second condition is only for convenience and, if relaxed,
the following conclusions remain almost unaltered.
}
The scaling dimensions coincide in the limit too, implying $\Delta_{\tilde{E}}|_{N=0}=\Delta_{E}|_{N=0}$,
therefore we argue on general grounds that there is a quantity $\Delta'_{E}$ such that
$\Delta_{\tilde{E}}-\Delta_{E}= N \Delta'_{E} +{\cal O}(N^2)$.
Notice that $\Delta'_{E}$ is a difference of the scaling dimensions, rather than a derivative, despite the notation.
It can be defined operatively as
\begin{equation}
 \Delta'_{E} \equiv \lim_{N\to 0} \frac{\Delta_{\tilde{E}}-\Delta_{E}}{N}\,.
\end{equation}
Now we insert the scaling limit of the correlators in the quenched averages and take the limit $N\to 0$ according
to our definitions. We find
\begin{equation}\label{eq:quenched-correlators-log}
\begin{split}
 &\overline{\langle E(x) E(0)\rangle} -\overline{\langle E(x) \rangle \langle E(0)\rangle}
 \sim  A(0)  \left|x\right|^{-2\Delta_{E}}  \,,\\
 &\overline{\langle E(x) \rangle \langle E(0)\rangle}
 \sim 2 A(0) \Delta'_{E}  \log\left|x\right|  \left|x\right|^{-2\Delta_{E}} \,,
\end{split}
\end{equation}
in which the scaling dimensions on the r.h.s.\ are implicitly evaluated at $N=0$
in order to not overburden the notation. From now on quantities will be understood at $N=0$, therefore
$\Delta_E|_{N=0}= \Delta_E$ and $\Delta_{\tilde{E}}|_{N=0}= \Delta_{\tilde{E}}$.

The surprising feature of \eqref{eq:quenched-correlators-log} is that the quenched average
of the disconnected part of the energy-energy correlation function displays a logarithmic behavior at criticality
in apparent violation of CFT basic properties!
What is happening here is that the operators are not eigenstates of the action of dilatation at (and only at) $N=0$.
More technically, the operators arrange in a two-by-two Jordan cell which is triangular and cannot be diagonalized
(in general cases with more degenerate operators the cell can have higher rank). Manipulations of the Jordan cell
result in a so-called logarithmic pair of operators \cite{Gurarie:1993, Flohr:2001zs, Gaberdiel:2001tr, Creutzig:2013hma, Gurarie:2013tma}.
An interesting conclusion, which is relevant to the renormalization group point of view,
comes from recalling that
the quantities $\Delta_{\tilde{E}}$ and $\Delta_{E}$ are universal functions of $N$,
therefore we have that $\Delta'_{E}$ is also universal.
Consequently, one is tempted to look for an observable to measure $\Delta'_{E}$
and the logarithmic behavior, which is easily found by measuring the ratio of the correlators at criticality
\begin{eqnarray}
 \frac{\overline{\langle E(x) \rangle \langle E(0)\rangle}}{\overline{\langle E(x) E(0)\rangle} -\overline{\langle E(x) \rangle \langle E(0)\rangle}}
 \sim  2 \Delta'_{E}  \log\left|x\right| \,.
\end{eqnarray}
Similar observables can be constructed for other logarithmic CFTs, including percolations,
loop-erased random walks, and self-avoiding walks. Logarithmic behavior is very common among CFTs
which can be described by some parameter, such as $N$, that takes a \emph{special} value through analytic continuation.

Having mentioned that the quantities $\Delta_{E}$, $\Delta_{\tilde{E}}$ and $\Delta'_{E}$ are universal,
it is clear that they can be computed by some renormalization group method. In fact, they are already known
to varying degree of accuracy and in several schemes for $\phi^4$-type theories, for example.
It is important to stress, however, that our discussion is certainly not limited to $\phi^4$-type theories of the Ising universality class, but potentially includes any possible quenched generalization of critical and multicritical
magnetic systems.
Furthermore, the quantity $\Delta'_{E}$ is in general much less known, if known at all.
A traditional starting point for any renormalization group based computation is
perturbation theory and the $\epsilon$-expansion,
which we adopt in the next sections to give several estimates and the above universal quantities.

\section{Renormalization group and critical properties}
\label{sec:method}

The truncation of the cumulant expansion \eqref{eq:cumulantsexp} to order $\kappa_n$ results in a model
with $\phi^{2n}$-type interactions, hence multicritical. Simple dimensional analysis shows that to the highest order
interactions correspond upper critical dimension $d_{c} = \frac{2n}{n-1}$.
For example, for $n=2$ the interaction is $\phi^{4}$-type, which has upper critical dimension $d_c=4$. More generally,
higher order interactions have smaller $d_c$ and accumulate to $d_c\to 2$ for increasing $n$.
These critical dimensions are the same as the ones of single-fields multicritical models, which are known to
interpolate to the multicritical CFTs of Zamolodchikov in $d=2$ \cite{zamolodchikov:1986}.

From our point of view, following ideas discussed in Refs.~\cite{Codello:2017a, Codello:2017qek, Codello:2019}, we take the possible values of $d_c$, and therefore the
label $n$, as an input for the classification of the multicritical generalizations. In fact,
at $d=d_c$ it is possible to construct a meaningful perturbative series in terms of the marginal couplings,
and at $d=d_c-\epsilon$ it is possible to determine universal quantities, such as critical exponents,
in the $\epsilon$-expansion. In the next sections we concentrate on the first few values $d_c=4,3,\frac{8}{3}$.
We also discuss a model with $d_c=\frac{10}{3}$, which we explain in more detail in the corresponding section.

For each $d_c$ there are very general forms of the renormalization group flows $\beta_V$ and $\beta_Z$,
respectively of the effective potential $V$ and of the wavefunction renormalization $Z$, which are determined
to a certain loop order. The amount of loops at which each multicritical RG flow is known varies considerably among the models, therefore we specify it in each corresponding section, however within our selection we consistently
have results between three and four loops.
In practice, the beta functional $\beta_V$ is used to determine the beta functions $\beta_i$ of the couplings,
in which the label $i$ runs among all couplings of $V$,
while $\beta_Z$ determines the anomalous dimension of the fields' multiplet.
The beta functions have increasing number of fixed points of order $\epsilon$ for increasing number
of couplings, so in the following we make a selection of the ones which we think are most important.
Using the fixed points, one can translate the perturbative series in the couplings into $\epsilon$-series.
The spectrum of the  relevant operators of the theory is given by the eigendeformations
of the stability matrix $M_{ij} \equiv \partial_j \beta_i$, whose eigenvalues $\theta_a$ are related to the scaling dimensions $\Delta_a$
of scaling operators (eigenvectors) by the relation $\theta_a = d - \Delta_a$.\footnote{%
Here we are referring to $\Delta_a$ as being the same as the CFT scaling dimension, notice however
that this relation is slightly modified if the corresponding operator is a CFT descendant \cite{Codello:2017a}.
This will not concern the results of this paper.
}

In dealing with the quadratic operators \eqref{eq:irreps}, we find it more convenient to renormalize them as composites.
To achieve this within a functional scheme, it is sufficient to operate the replacement
$V \to V + J_R\cdot R$ in $\beta_V$, which gives the RG running of the composite operators
if $R=S,X,Y$ runs through the irreducible representations and $J_R$ are sources with the correct number of indices
and symmetries. We used the linear order in $J_R$ of the substitution, to determine the critical exponents and scaling dimensions of $R$, but in principle one could go beyond and compute the operator product expansion
of pairs of operators \cite{Pagani:2020ejb}.

The first most important field-theoretical critical exponent is the anomalous dimension $\eta$,
which is obtained by diagonalizing $\beta_Z$. In all even models, the symmetry $H_N$ is enough to
constrain all fields to have the same anomalous dimension.
The second most important field-theoretical critical exponent is $\nu$, that controls the scaling
of the correlation length and is estimated from the inverse of the quadratic singlet in $V$.
From \eqref{eq:irreps}, we expect three different exponents for general $N$, and correspondingly three different
inverse quantities that we denote $\nu_R$ for $R=S,X,Y$.
Obviously, $\nu=\nu_S$ which we can verify easily for all computations.
As discussed in Sect.~\ref{sec:logarithms}, we have that $\nu_S=\nu_X$ while $\nu_S\neq \nu_Y $ in the limit $N\to 0$.
This has an interesting consequence as we are about to see.

As discussed in Ref.~\cite{Wiese:2020}, the critical exponent of the operator $X$ of \eqref{eq:irreps}
is also the fractal dimension of the propagator lines of the model, therefore
$ d_f = \frac{1}{\nu_X}$. Using again the fact that the operators $S$ and $X$ degenerate in the limit,
we can establish that
\begin{equation}
 \begin{split}
 d_f= \frac{1}{\nu}
 \end{split}
\end{equation}
for all the random models that we are about to discuss.
Notice that this result replicates similar statements found in Ref.~\cite{Wiese:2020} in the same limit, but for different
symmetry content.
In fact, in that paper the limit $N\to 0$ of the $O(N)$ model is considered, but, since $O(N)$
is the maximal symmetry of a model with $N$ scalars \cite{Osborn:2017ucf} and since $H_N\subset O(N)$,
our equations actually contain the case $O(N)$ as special case and the limits $N\to 0$ are actually the same.
The limit $N\to 0$ of the $O(N)$ model is interesting because it corresponds to the continuum
theory behind a self-avoiding walk, so a random walker which cannot cross its path (or, in other words,
whose line has exactly $N=0$ loops) \cite{DeGennes1972}.

The operators $X$ and $Y$ of \eqref{eq:irreps} are also interesting when discussing the breaking of $H_N$ symmetry.
One can imagine a situation in which the model is at the critical temperature, but it is also deformed by some component
of either $X$ or $Y$, which breaks explicitly $H_N$ (therefore the correlation length is still finite).
Using scaling analysis, it is easy to show (see again Ref.~\cite{Wiese:2020} for the $O(N)$ example)
that, depending on the irreducible deformation,
the model is crossing-over to a phase with a smaller symmetry group, $H_N\to G$,
and that the crossover is critical for small coupling, which means close to the original critical point.
The exponents controlling these transformations, which go by the name of \emph{crossover exponents},
are estimated as the ratios of $\nu$s
\begin{equation}
 \begin{split}
 \phi_X = \nu/\nu_X \,, \qquad\qquad\qquad \phi_Y = \nu /\nu_Y\,.
 \end{split}
\end{equation}
As before, we notice that for $N\to 0$ we have that $\phi_X=1$, which applies to all models described in the following
and therefore will not be repeated.
Instead, $\phi_Y $ is an independent exponent, which we give explicitly later even though it can easily be
derived from the operator scaling dimensions $\phi_Y  =\frac{d-\Delta_Y}{d-\Delta_S}$.
The quadratic deformation $Y$ induces a crossover to a (broken symmetry) phase described by the so-called Klein four-group $G = \mathbb{Z}_2 \rtimes S_2 \simeq \mathbb{K}_4 $ and which could be associated to the critical content of two coupled Ising models.

The final quantities that we compute are rooted in both RG and CFT formalisms. The $A$-function is
the scalar function from which one can derive the RG flow as a gradient for all the even models. If $g_a$ are all the couplings and $\beta_a$ the corresponding beta functions, then $A$ is derived implicitly from
\begin{equation}
 \begin{split}
 \beta_a = \sum_b h_{ab} \frac{\partial A}{\partial g_b} \,.
 \end{split}
\end{equation}
In the rest of the paper we assume that the metric $h_{ab}$ in the space of the couplings is flat,
which is consistent until the next-to-leading order.\footnote{%
To be precise, the metric is flat only for a specific choice of couplings.
In the case $d_c=4$, the most general quartic potential
is $V(\phi)= \sum\lambda_{ijkl}\phi_i\phi_j\phi_k\phi_l$, with couplings $\lambda_{ijkl}$ which are fully symmetric tensors. Their beta functions $\beta_{ijkl}$ can be derived from the function $A$ as $\beta_{ijkl}= \frac{\partial A}{\partial \lambda_{ijkl}}$, thus we choose the metric to be flat in the coordinates $\lambda_{ijkl}$. If we were to specialize
the same formula to the couplings of \eqref{eq:replicated-action}, we would generally find a nontrivial metric \cite{Osborn:2017ucf}.
}
Naively, $A$ counts the number of degrees of freedom of the model. Since $N\to 0$ corresponds to a situation
with no fields, we have that in our cases $A \sim N$, implying that the quantity of interest is instead
\begin{equation}
 \begin{split}
 a \equiv \lim_{N\to0} \frac{A}{N}\,,
 \end{split}
\end{equation}
for which we stress the similarity with $\Delta'_E$ of  Sect.~\ref{sec:logarithms}.
Finally, the constant $C_T$ appears when expanding the correlator involving the trace of the stress-energy tensor, $T$,
and two copies of the field $\phi_i$ for $d=4$. Loosely speaking
\begin{equation}
 \begin{split}
  C_T \sim \langle \phi_i \phi_i T\rangle\,,
 \end{split}
\end{equation}
as shown in Ref.~\cite{Dey:2016}, in which the relation with the marginal couplings of $V$ has also been derived.
We have that  $C_T \sim N$, likewise $A$, so again the interesting result comes from the limit
\begin{equation}
 \begin{split}
 c_T \equiv \lim_{N\to0} \frac{C_T}{N}\,.
 \end{split}
\end{equation}
For both $a$ and $c_T$, we can think at their definitions as being the uppercase quantities
normalized with their free theory counterparts
\begin{equation}
 \begin{split}
 a \equiv \lim_{N\to0} \frac{A}{A_{\rm free}}\,, \qquad \qquad c_T \equiv \lim_{N\to0} \frac{C_T}{C_{T\, {\rm free}}}\,.
 \end{split}
\end{equation}
Of course, the free theory that we refer to here has $N$ free scalar fields,
and not just one as in the normalization given in Ref.~\cite{Dey:2016}.
We finally notice that at LO the following relation is true
\begin{equation}
  A= -\frac{3 \epsilon}{2} \sum_a \eta_a =- \frac{3 \epsilon}{2}N \eta \,,
\end{equation}
from which it directly follows that $a = -\frac{3\epsilon}{2} \eta$.

\section{\texorpdfstring{Tricritical theory in $d=3-\epsilon$}{Tricritical theory in 3-epsilon}}
\label{sec:dc3}

The renormalization group analysis in $d=3-\epsilon$ dimensions determines the properties of tricritical fixed points \cite{Stephen1973, Lewis1978}.
This is certainly true for a single component field $\phi$ with $\mathbb{Z}_2$-symmetry,
for which there exists a tricritical point which
requires the simultaneous fine-tuning at criticality of two relevant operators, besides
the external magnetic field, hence the name ``tricritical''.
In the generalization to $H_N$ symmetry and the randomly diluted case, however, the fixed points emerging in $d=3-\epsilon$ dimensions might require the fine-tuning at criticality of more than two parameters and, strictly speaking, one should regard them as multicritical ones. That being said, following \cite{Stephen1976} and in order to make clear contact
with the corresponding pure theory, we refer to the fixed points emerging at $d_c=3$ in the limit $N\to 0$
as tricritical ones.

Truncating the cumulant expansion at third order, we are left with the following marginal potential
\begin{equation}\label{eq:marginalV3}
\begin{split}
  V(\phi_a)  &=
 w_1 \sum_a E_a^3(x)  + w_2 \sum_{a \neq b} E_a^2(x)E_b(x) + w_3 \sum_{a\neq b \neq c} E_a(x) E_b(x) E_c(x) \\
 &=
 \upsilon_1\Bigl(\sum_a \phi_a^2\Bigr)^3 + \upsilon_2 \sum_a \phi_a^2 \sum_b \phi_b^4 +  \upsilon_3 \sum_a \phi_a^6
\,.
\end{split}
\end{equation}
We have introduced a second parametrization that makes explicit the form of the interaction
in terms of the basic fields and shows that the $O(N)$ limit is reached when $\upsilon_2=\upsilon_3=0$.
The linear relations between the couplings $\upsilon_i$ and $w_i$ are trivial to obtain and we only need $w_1=(\upsilon_1+\upsilon_2+\upsilon_3)$ in the following.

We computed the beta functions of the RG flow for \eqref{eq:marginalV3}; they can be found in \eqref{eq:betasystem-dc3}.
Apart for the trivial Gaussian fixed point, the theory admits the tricritical pure Ising fixed point \cite{Lewis1978} with coordinates
$\{  \upsilon_1^{\star} = 0 \,,  \upsilon_2^{\star} = 0 \,, \upsilon_3^{\star} = \frac{3}{10}\epsilon  \}$
and LO anomalous dimension given by $\eta = \epsilon^2/500$.
Then there is the tricritical $O(N)$ fixed point in the limit $N\to 0$ with coordinates
$\{ \upsilon_1^{\star} = \frac{15}{44} \epsilon \,, \upsilon_2^{\star} = 0 \,, \upsilon_3^{\star} = 0 \,, \}$, which
corresponds to the tricritical SAW universality class and has anomalous dimension given by $\eta = \epsilon^2/726$;
it has been linked to the so-called Flory $\theta$-point of very long polymer chains \cite{DeGennes1972,DeGennes1975, Duplantier1982}.

Finally, there are three further fixed points that display genuine $H_N$ symmetry in the limit $N\to 0$;
therefore, there are three potentially equivalent candidates for random tricritical model.
Using the experience on the standard critical case, which is reviewed in App.~\ref{sec:dc4}, we apply a rationale that boils this list down to a single interesting fixed point.
Since we are interested in phase transitions that can be of second order in the pure system, a first physical requirement is that of stability, which is equivalent to
$w_1^\star = (\upsilon_1^\star+\upsilon_2^\star+\upsilon_3^\star)>0$
that ensures that \eqref{eq:marginalV3} is bounded from below
(an analog condition holds for the case $d_c=4$).
A second physical requirement is that the theory is perturbatively unitary:
this implies that the anomalous dimension $\eta$ must be positive\footnote{%
The anomalous dimension is \emph{not} perturbatively positive for the standard diluted model in $d=4-\epsilon$, as shown in App.~\ref{sec:dc4}. However, it \emph{is} positive for $d=3$,
implying that it must change sign through a nonperturbative mechanism. This is very important for the comparison with lattice simulations, which clearly display $\eta>0$.
We elect the positivity of the anomalous dimension as a physical requirement for this reason.
}
and that the spectrum of RG deformations is \emph{real}.

At NLO, corresponding to four loops, this leaves us only with the following FP
\begin{equation}
\begin{split}
  \upsilon_1^{\star} & = 0.427688 ~\epsilon + 3.6999 ~\epsilon ^2 \,,\\
  \upsilon_2^{\star} & = -0.164358 ~\epsilon -2.49894 ~\epsilon ^2 \,, \\
  \upsilon_3^{\star} & =  0.084626 ~\epsilon +1.26481 ~\epsilon ^2  \,,
\end{split}
\end{equation}
The above solution can be determined analytically, and at the leading order
it comes from solving three quadratic equations in three variables; the solution, however,
is enormous, so we prefer to report it numerically.

The critical exponents $\eta$ and $\nu$ for this fixed point are given by
\begin{equation}
\begin{split}
  \eta  =  0.00137842 ~\epsilon^2\,, \qquad \qquad
  \nu  = \frac{1}{2} + 0.00551369 ~\epsilon ^2 \,,
\end{split}
\end{equation}
while the scaling dimensions associated to the irreps.\ \eqref{eq:irreps} read
\begin{equation}
\begin{split}
  \Delta_E & = 1 - \epsilon + 0.0220548 ~\epsilon ^2 \,,\qquad \qquad
  \Delta_Y = 1 - \epsilon + 0.0222901 ~\epsilon ^2 \,,\\
  \Delta'_E & = -0.00832891 ~\epsilon ^2 \,.
\end{split}
\end{equation}
The $a$-function in this case takes the negative value
\begin{equation}
  a^\star = -0.0206763 ~\epsilon ^3.
\end{equation}
We conclude giving the fractal dimension of propagator lines and the nontrivial crossover exponent
\begin{equation}
\begin{split}
  d_f  = 2 - 0.0220548 ~\epsilon ^2 \,, \qquad \qquad
  \phi_Y  = 1 - 0.000117671 ~\epsilon ^2\,.
\end{split}
\end{equation}

To conclude, we briefly discuss some property of the two other fixed points of \eqref{eq:marginalV3} that have been left out of the discussion.
One of them lies in the unphysical region $w_1<0$ and thus is not positive definite.
The other one has a complex conjugate pair of eigenvalues in the stability matrix (and therefore has \emph{complex} spectrum).
We discarded both of them accordingly, even though they might have some physical application.
For another analysis of the random tricritical model we refer to \cite{Stephen1976}.

\section{\texorpdfstring{Tetracritical theory in $d=\frac{8}{3}-\epsilon$}{Tetracritical theory in 8/3-epsilon}}
\label{sec:dc8/3}

The next in line among the possible generalizations comes from the $\epsilon$-expansion below $d_c=\frac{8}{3}$,
which determines the properties of tetracritical fixed points.
We proceed by truncating the cumulant expansion to the fourth order, obtaining the following marginal potential
\begin{eqnarray}\label{eq:marginalV8/3}
  V(\phi_a) & =& u_1 \sum_a E_a^4(x) + u_2 \sum_{a \neq b} E_a^2(x)E_b^2(x) + u_3 \sum_{a \neq b}    E_a^3(x) E_b(x)
 \nonumber\\
            && + u_4 \sum_{a \neq b \neq c} E^2_a(x)E_b(x)E_c(x) + u_5 \sum_{a \neq b \neq c \neq l} E_a(x) E_b(x) E_c(x) E_l(x) \\
 &=& \rho_1 \Bigl(\sum_a \phi_a^2\Bigr)^4 + \rho_2 \Bigl(\sum_a \phi_a^2\Bigr)^2 \sum_b \phi_b^4
 + \rho_3 \sum_a \phi_a^2 \sum_b \phi_b^6 + \rho_4 \Bigl(\sum_a \phi_a^4 \Bigr)^2
 + \rho_5 \sum_a \phi_a^8
\nonumber
\,.
\end{eqnarray}
As in the previous sections the second parametrization of couplings $\rho_i$ conveniently
highlights the $O(N)$ limit as the case in which only $\rho_1\neq 0$.

The beta functions of the RG flow for \eqref{eq:marginalV8/3} are given in \eqref{eq:betasystem-dc8/3}.
The theory admits several nontrivial fixed points. First, we have the tetracritical pure Ising fixed point with coordinates
$\{\rho_1^{\star} = 0 \,, ~ \rho_2^{\star} = 0 \,, ~ \rho_3^{\star} = 0 \,, ~ \rho_4^{\star}  = 0 \,, ~ \rho_5^{\star}  = \frac{3}{70} \epsilon \}$,
and LO anomalous dimension given by $\eta = \frac{9}{85750} \epsilon^2$, which can be checked against Ref.~\cite{Zinati:2019gct}.
Then, we have the tetracritical $O(N)$ fixed point in the limit $N\to 0$ with coordinates $\{ \rho_1^{\star}  = \frac{105}{2144} \epsilon\,,
 ~\rho_2^{\star}  = 0 \,,  ~ \rho_3^{\star} = 0 \,, ~ \rho_4^{\star} = 0 \,, ~ \rho_5^{\star} = 0 \}$
and anomalous dimension $\eta = \frac{9}{143648} \epsilon^2$,
corresponding to the tetracritical SAW universality.

We then have several nontrivial fixed points with true hypercubic symmetry.
The requirement that $u_1 ^{\star} \equiv ( \rho_1^\star + \rho_2^\star + \rho_3^\star + \rho_4^\star + \rho_5^\star)>0$
leaves us with seven possible candidates.
Among these, only three have a positive anomalous dimension $\eta$, and only one has a completely real spectrum.
These features are in common with the random Ising model in $d=3$
(see the appropriate comment in Sect.~\ref{sec:dc4}) and with the tricritical example of the previous section. The corresponding LO fixed point coordinates are determined at three loops as
\begin{equation}
  \{ \rho_1^{\star}  =  0.009405 ~\epsilon \,,  ~ \rho_2^{\star} = 0.060199 ~\epsilon \,, ~ \rho_3^{\star} = 0.002977 ~\epsilon\,, ~ \rho_4^{\star} = 0.022737 ~\epsilon \,, ~ \rho_5^{\star}  = -0.048088 ~\epsilon \} \,.
\end{equation}
The anomalous dimension is given by $\eta = 0.0000615196 \epsilon^2$. In principle,
even the two discarded fixed points may as well be interesting;
they have a complex conjugate pair of critical exponents in the stability matrix, but positive anomalous dimension.
We chose to limit our discussion to fixed points with real spectrum for clarity,
assuming that it is a physically meaningful requirement based on the analogy with the previous models.

\section{\texorpdfstring{Multicritical theory in $d=\frac{10}{3}-\epsilon$}{Multicritical theory in 10/3-epsilon}}
\label{sec:dc10/3}

In the previous sections, we exclusively considered even interactions of the fields
and each model can be regarded as a generalization of the tower of multicritical models $\phi^{2n}$
by Zamolodchikov \cite{zamolodchikov:1986, itzykson1991sft, ODwyer:2007}.
A way around this limitation is to include a field singlet $\sigma$ in the multiplet of fields,
$\phi_a \to (\phi_a,\sigma)$.
The presence of a singlet allows us to construct interactions with an odd number of fields.
For the single component case, these interactions
also correspond to a tower of multicritical models $\phi^{2n+1}$,
albeit much less known than the even counterpart \cite{Zambelli:2016,Codello:2017}.
The simplest possible odd theory with $H_N$ symmetry
is the one with a cubic interaction and upper critical dimension $d_c=6$,
however the potential is constrained to be $V\sim \sum_a\sigma (\phi_a)^2$, and therefore it actually has
enhanced $O(N)$ symmetry. This is essentially the same theory considered in Ref.~\cite{Giombi:2019},
when attempting to construct the Wilson-Fisher $O(N)$ model above the upper critical dimension $d_c=4$.
The result is a non-unitary theory as clarified recently in \cite{Giombi:2019}.

For our construction, we necessitate a true symmetry $H_N$, which can be achieved by
considering at least a theory with quintic interaction and upper critical dimension $d_c=\frac{10}{3}$.
The critical potential is
\begin{equation}\label{eq:marginalV103}
\begin{split}
  V(\phi_a,\sigma) &= v_1 \sigma^5 + v_2 \sigma^3 \sum_a E_a + v_3 \sigma\sum_{a\neq b}E_a E_b
 + v_4 \sigma\sum_a E_a^2 \\
 &=
 \kappa_1 \sigma^5 + \kappa_2 \sigma^3 \sum_a \phi_a^2 + \kappa_3 \sigma \Bigl(\sum_a \phi_a^2\Bigr)^2 + \kappa_4 \sigma \sum_a \phi_a^4
 \,,
\end{split}
\end{equation}
and it can accommodate a true $H_N$ symmetry thanks to the monomial coupled to $v_3$ in the first parametrization
or, equivalently, $\kappa_4$ in the second.
For this reason, a nonzero $v_3$ is a signature of genuine hypercubic symmetry,
instead of the larger group $O(N)$.
If we were to follow the same logic of the construction of Ref.~\cite{Giombi:2019},
then the resulting theory could be interpreted as an attempt to promote the
model with hypercubic symmetry and $d_c=3$ above its upper critical dimension.

For general $N$, this model generalizes the single-field case presented in detail in Ref.~\cite{Codello:2017}.
The corresponding RG flow can be read off from the equations presented in Ref.~\cite{Codello:2017}
by opportunely adding the field's indices, which can be done in a unique way at the leading order.
We also adopt the same normalization to avoid factors of $4\pi$.
The result of the RG analysis is that there are several fixed points, most of which are genuinely
complex, having nonzero real and imaginary parts.
As in the single field case, however,
there are some purely imaginary solutions. In fact, purely imaginary solutions are protected
by the parity and time-reversal pseudo-symmetry
$$
{\cal PT}: V(\phi_a,\sigma) \to V^*(\phi_a,-\sigma)\,,
$$
in which we denoted complex-conjugation of the potential with the asterisk.
The same happens for the Lee-Yang model \cite{Bender:2012ea, Bender:2013qp}.
We deduce that the resulting models are non-unitary, but they are still interesting
because the boundedness of the spectrum is guaranteed by the action of ${\cal PT}$.
Non-unitarity results in $\eta<0$, so for this section we relax the requirement of positive
anomalous dimension.

An interesting aspect of the analysis of the model with odd interactions
is related to the generalization of the quadratic composite operators \eqref{eq:irreps}.
Having a new singlet field $\sigma$ at our disposal, it is clear that it is possible to add a new
quadratic singlet to \eqref{eq:irreps}, which we denote $S'=\sigma^2$.
In general, the renormalization process will mix the singlets $S$ and $S'$ because they carry the same
labels for arbitrary values of $N$.
However, this notion clashes with the statement that the singlet $S$ and the tensor $X$ have degenerate scaling dimensions on the basis of $H_N$ symmetry in the limit $N\to 0$,
which was verified multiple times in the previous sections.
This conundrum is solved by noticing that, precisely in the limit $N\to 0$, the mixing matrix of
$S$ and $S'$ is diagonal and they become separate scaling operators,
which is thus consistent with $S$ alone becoming degenerate with $X$.
In the limit $N\to 0$, we thus have two distinct critical exponents $\nu_\phi$ and $\nu_\sigma$
from $S$ and $S'$ respectively,
as well as two distinct anomalous dimensions $\eta_\phi$ and $\eta_\sigma$.

The RG flow of \eqref{eq:marginalV103} is given in \eqref{eq:betasystem-dc10/3}.
We find two fixed points that fit all the criteria and therefore we think are interesting to report.
The coordinates of the first fixed point are
\begin{equation*}
  \{ \kappa_1^\star = \frac{3}{\sqrt{229}} \,i ~\epsilon^{1/2} ,
~ \kappa_2^\star = 0.948238 \,i ~\epsilon^{1/2},
~ \kappa_3^\star =   -0.630458 \,i ~\epsilon^{1/2},
~ \kappa_4^\star = 0.775543 ~ \,i ~\epsilon^{1/2} \} \,,
\end{equation*}
and its associated critical exponents are
\begin{align}
  \eta_\phi & = -0.00048373 ~\epsilon \,,   &\nu_\phi & = \frac{1}{2} + 0.00196926 ~\epsilon \,, \\
  \eta_\sigma & = -\frac{3}{1145} ~\epsilon \,, &\nu_\sigma & = \frac{1}{2}- \frac{33}{4580} ~\epsilon \,,
\end{align}
The coordinates of the second fixed point are
\begin{equation*}
  \{ \kappa_1^\star = \frac{3}{\sqrt{229}} \,i ~\epsilon^{1/2} ,
~ \kappa_2^\star = -0.390662 \,i ~\epsilon^{1/2} ,
 ~ \kappa_3^\star = -0.575971 \,i ~\epsilon^{1/2} ,
~ \kappa_4^\star = -0.0131081 \,i ~\epsilon^{1/2} \} \,,
\end{equation*}
and the corresponding critical exponents are
\begin{align}
  \eta_\phi & = -0.00410786  ~\epsilon \,,   &\nu_\phi & = \frac{1}{2} - 0.00822047 ~\epsilon \,, \\
  \eta_\sigma & =-\frac{3}{1145} ~\epsilon \,, &\nu_\sigma & = \frac{1}{2}- \frac{33}{4580} ~\epsilon \,.
\end{align}
We notice that the critical exponents
associated to the singlet $\sigma$ are identical at the two fixed points and could be given as simple fractions.
This happens through the combination of few facts: first, the beta function of $v_1$ is decoupled from the rest of the system in the limit $N\to 0$, resulting in $\beta_{\kappa_1}=-\frac{3}{2}\epsilon\, \kappa_1 -\frac{229}{6}(\kappa_1)^3$.
Second, both the anomalous dimension $\eta_\sigma=\frac{(\kappa_1)^2}{15}$ and the scaling exponent $\frac{1}{\nu_\sigma}=2-\frac{11}{15}(\kappa_1)^2$ of $\sigma^2$ only depend on $\kappa_1$ in the limit.

\section{Conclusions}
\label{sec:conclusions}

There are two main messages that we wanted to deliver through this paper:
the first one is that there are several, possibly infinitely many, multicritical generalizations of the hypercubic model.
These generalizations are relevant if one desires to approach the problem
of describing quenched averages of pure magnetic systems with arbitrary
criticality order or arbitrary distributions of the disorder.
We have argued in favor of the existence of such generalizations after a summary
of the replica approach to describe disorder, in which we evidenced the role that a nontrivial distribution
of the disorder has in generating higher order terms for the replicated action.
While most of the previous literature has concentrated attention on a Gaussian noise,
eventually invoking universality to justify that all distributions should essentially fall into the same universality class,
we believe that a sufficiently complex distribution might trigger a multicritical behavior in the random system.
If this is correct, our findings suggest that such type of multicritical behaviors can be classified.
For this purpose, we have used perturbation theory and the $\epsilon$-expansion as classification tools,
determining the upper critical dimension indirectly from the choice of the marginal interactions that control
the perturbative series.
Given that almost all the new upper critical dimensions that we have studied are below $d=3$,
we also believe that our findings might be most relevant for systems with two physical dimensions
and therefore could have interesting implications in the context of low-dimensional physics and
two-dimensional CFT.

The second main message is that some conformal data and the scaling of very interesting observables are accessible by renormalization group methods, including less-known universal coefficients that are related to logarithmic corrections to CFT. These corrections are not logarithmic in the sense of mean-field, but instead literally display logarithms as part of the CFT correlators.
We have discussed in few simple steps how to access some quantities that are relevant for the study of CFTs
and log-CFTs by a combination of group representation theory and renormalization group methods making an explicit connection with earlier work by Cardy.
Our hope is to open a new avenue for comparing and combining results coming from RG and CFT,
since both methods might have advantages and disadvantages.
For example, it is very simple to handle the analytic continuation
in the number $N$ of replicas through the methods of this paper.
An interesting aspect for a generalization, in this regard,
would be to promote the functional approach to a nonperturbative method of the RG,
such as the Wetterich equation \cite{Wetterich:1992yh}.
How to properly continue to $N\to 0$ the fully functional form
of the local potential approximation is, however, presently unknown to us,
as it would imply to have a function with zero arguments.
It is still possible to continue the couplings' beta functions as done
in Ref.~\cite{PhysRevB.65.140402} for the hypercubic model
and in Ref.~\cite{Zinati:2017hdy} for the comparable Potts model.

In passing, we have mentioned several other fixed points and relative critical models, besides the limit $N\to 0$
of the hypercubic.
Our hope is to have mentioned a complete accounting of the interesting fixed points
appearing in the limit of ``zero fields.''
Among these, we have seen several multicritical generalizations
of the self avoiding walk, which arise as the limit $N\to 0$ of models with $O(N)$ symmetry.
This is an important point because, especially for some of models that we have considered,
the RG diagram contains many new fixed points
which are either complex (thus maybe related to some complex CFT) or non-unitary
(in the sense that they do not satisfy the unitarity bound perturbatively)
and some rationale must be used to infer which are the important ones.
For brevity, we have made a selection of the interesting candidate fixed points
to generalize the hypercubic behavior in the limit and decided to not dive too deeply
in the study of all others. It is still possible, however,
that the points that we have not discussed in detail have some interesting physical meaning,
so we hope that our work triggers some interest in their respect as well.

\paragraph{Acknowledgements.}
We are grateful to A.~Stergiou for suggesting us some important references
and to E.~Vicari for pointing our interest to the $N\to 0$ limit of the hypercubic model.
For performing the computations of this paper we relied heavily on the \emph{Mathematica} packages \cite{xact-package,xperm-package} and \cite{Nutma:2013zea}. RBAZ acknowledges the support from the
French ANR through the project NeqFluids (grant ANR-
18-CE92-0019).\enlargethispage{\baselineskip}

\appendix

\section{Beta functions}
\label{app:betas}

For each upper critical dimensions $d_c$, we report the diagrammatic expressions
of the beta functional of the potential $\beta_V$ and of the anomalous dimension
matrix $\gamma_{ij}$ (the latter is given in the form of the singlet $\gamma_{ij}\phi_i\phi_j$
for compactness).
We also give the explicit expressions of the $\beta$-functions,
that can be obtained by ``simply'' inserting the explicit form
of the potentials and iteratively simplifying the algebra of generalized Kronecker symbols
that are responsible for the contractions of the fields in $V(\phi)$.
The solutions of the $\beta$-systems are the fixed-point coordinates given in the main text.
All critical exponents and scaling dimensions can be extracted from the same beta functionals:
as explained in Sect.~\ref{sec:method}, it is sufficient to perform the replacement
$V \to V + J_R\cdot R$ in $\beta_V$ for some scaling relevant composite operator $R$
and linearize the functional flow. We do not show any $\gamma$-function,
because the $\epsilon$-expansions of all interesting critical exponents are given in the main text.

For displaying the beta functionals, we use a diagrammatic representation
in which $n$-vertices correspond to $n$ derivatives of the potential w.r.t.\ the fields
\begin{equation}
 \begin{split}
   V_{i_1 \,\cdots \, i_n} \equiv \frac{\partial^n V}{\partial\phi_{i_1} \cdots \partial \phi_{i_n}}
  = \begin{tikzpicture}[baseline=-.1cm]
      \draw (1,.5) to[out=180,in=90] (0,-.5);
      \draw (1,.5) to[out=180+20,in=90] (.4,-.5);
      \draw (1,.5) to[out=180+52,in=90] (.8,-.5);
      \draw (1,.5) to[out=0,in=90] (2,-.5);
      \filldraw [gray!50] (1,.5) circle (2pt);
      \draw(1,.5) circle (2pt);
      \filldraw [white!50] (0,-.5) circle (2pt);
      \draw(0,-.5) circle (2pt);
      \filldraw [white!50] (.4,-.5) circle (2pt);
      \draw(.4,-.5) circle (2pt);
      \filldraw [white!50] (.8,-.5) circle (2pt);
      \draw(.8,-.5) circle (2pt);
      \filldraw [white!50] (2,-.5) circle (2pt);
      \draw(2,-.5) circle (2pt);
       \draw[] (0,-.5) node[below] { ${\scriptstyle i_1}$};
       \draw[] (.4,-.5) node[below] { ${\scriptstyle i_2}$};
       \draw[] (.8,-.5) node[below] { ${\scriptstyle i_3}$};
       \draw[] (2,-.5) node[below] { ${\scriptstyle i_n}$};
       \draw[] (1.4,-.5) node[below] { ${\scriptstyle \cdots}$};
      \end{tikzpicture}
  \,,
 \end{split}
\end{equation}
in which the gray circle represents the $n$-vertex in question,
hollow circles are the other vertices,
and lines organize the contractions of the indices.
For example, the following diagram represents
\begin{equation}
\begin{split}
        \begin{tikzpicture}[baseline=-.1cm]
      \draw (1.5,0) circle (.5cm);
      \draw (1,0) to[out=-45,in=135] (1.5,-.5);
      \filldraw [gray!50] (1.5,.5) circle (2pt);
      \draw(1.5,.5) circle (2pt);
      \filldraw [gray!50] (2,0) circle (2pt);
      \draw (2,0) circle (2pt);
      \draw (1,0) to[out=0,in=90] (1.5,-.5);
      \filldraw [gray!50] (1,0) circle (2pt);
      \draw (1,0) circle (2pt);
      \filldraw [gray!50] (1.5,-.5) circle (2pt);
      \draw(1.5,-.5) circle (2pt);
      \end{tikzpicture}
& = V_{ij}V_{jklm}V_{klmp}V_{pi}\,,
\end{split}
\end{equation}
where we started labelling from the top vertex and moved clockwise.
These \emph{are not} Feynman diagrams,
but the loop-counting that they display is the correct one.

\subsection{\texorpdfstring{\boldmath{$d_c=4$}}{beta dc=4}}

The RG flow of a theory with quartic interaction in $d=4-\epsilon$
can be found in Ref.~\cite{Osborn:2017ucf}
at three loops in fully functional form, however the beta functions of the couplings are known to a much higher loop count (see \cite{Pelissetto:2007gw} and references therein).
\begin{equation}
\begin{split}
        \beta_V  = &
        + \frac{1}{2} ~
        \begin{tikzpicture}[baseline=-.1cm]
        \draw (0,0) circle (.5cm);
        \filldraw [gray!50] (.5,0) circle (2pt);
        \draw (.5,0) circle (2pt);
        \filldraw [gray!50] (-.5,0) circle (2pt);
        \draw (-.5,0) circle (2pt);
        \end{tikzpicture}
        ~ -\frac{1}{2} ~
        \begin{tikzpicture}[baseline=-.1cm]
        \draw (1.5,0) circle (.5cm);
        \draw (1,0) to[out=0,in=180] (2,0);
        \filldraw [gray!50] (1.5,.5) circle (2pt);
        \draw(1.5,.5) circle (2pt);
        \filldraw [gray!50] (1,0) circle (2pt);
        \draw (1,0) circle (2pt);
        \filldraw [gray!50] (2,0) circle (2pt);
        \draw (2,0) circle (2pt);
        \end{tikzpicture}
        ~ -\hspace{2pt}\frac{1}{8} \hspace{2pt} ~
        \begin{tikzpicture}[baseline=-.1cm]
        \draw (3,0) circle (.5cm);
        \draw (3,.5) to[out=-90,in=180] (3.5,0);
        \filldraw [gray!50] (3,.5) circle (2pt);
        \draw(3,.5) circle (2pt);
        \filldraw [gray!50] (3.5,0) circle (2pt);
        \draw (3.5,0) circle (2pt);
        \draw (2.5,0) to[out=0,in=90] (3,-.5);
        \filldraw [gray!50] (2.5,0) circle (2pt);
        \draw (2.5,0) circle (2pt);
        \filldraw [gray!50] (3,-.5) circle (2pt);
        \draw(3,-.5) circle (2pt);
        \end{tikzpicture}
        ~ +\frac{\zeta_3}{2} ~
        \begin{tikzpicture}[baseline=-.1cm]
        \draw (4.5,0) circle (.5cm);
        \draw (4.5,0) to [out=90,in=-90] (4.5,.5);
        \draw (4.5,0) to[out=-30,in=150] (4.969,-.171);
        \draw (4.5,0) to[out=210,in=30] (4.031,-.171);
        \filldraw [gray!50] (4.5,.5) circle (2pt);
        \draw (4.5,.5) circle (2pt);
        \filldraw [gray!50] (4.5,0) circle (2pt);
        \draw (4.5,0) circle (2pt);
        \filldraw [gray!50] (4.969,-.171) circle (2pt);
        \draw (4.969,-.171) circle (2pt);
        \filldraw [gray!50] (4.031,-.171) circle (2pt);
        \draw (4.031,-.171) circle (2pt);
        \end{tikzpicture}\\
        & -\frac{1}{4} ~
        \begin{tikzpicture}[baseline=-.1cm]
        \draw (6,0) circle (.5cm);
        \draw (6.5,0) to[out=180,in=90] (6,-.5);
        \filldraw [gray!50] (6,.5) circle (2pt);
        \draw(6,.5) circle (2pt);
        \filldraw [gray!50] (6.5,0) circle (2pt);
        \draw (6.5,0) circle (2pt);
        \draw (5.5,0) to[out=0,in=90] (6,-.5);
        \filldraw [gray!50] (5.5,0) circle (2pt);
        \draw (5.5,0) circle (2pt);
        \filldraw [gray!50] (6,-.5) circle (2pt);
        \draw(6,-.5) circle (2pt);
        \end{tikzpicture}
        ~ + \hspace{1pt} 2 \, ~
       \begin{tikzpicture}[baseline=-.1cm]
       \draw (0,0) circle (.5cm);
       \draw (-.5,0) to[out=0,in=180] (.5,0);
       \filldraw [gray!50] (0,.5) circle (2pt);
       \draw(0,.5) circle (2pt);
       \filldraw [gray!50] (.5,-0) circle (2pt);
       \draw (.5,0) circle (2pt);
       \draw (-.5,0) to[out=0,in=90] (0,-.5);
       \filldraw [gray!50] (-.5,0) circle (2pt);
       \draw (-.5,0) circle (2pt);
       \filldraw [gray!50] (0,-.5) circle (2pt);
       \draw(0,-.5) circle (2pt);
       \end{tikzpicture}
       \,-\frac{3}{16} ~
      \begin{tikzpicture}[baseline=-.1cm]
      \draw (1.5,0) circle (.5cm);
      \draw (1,0) to[out=-45,in=135] (1.5,-.5);
      \filldraw [gray!50] (1.5,.5) circle (2pt);
      \draw(1.5,.5) circle (2pt);
      \filldraw [gray!50] (2,0) circle (2pt);
      \draw (2,0) circle (2pt);
      \draw (1,0) to[out=0,in=90] (1.5,-.5);
      \filldraw [gray!50] (1,0) circle (2pt);
      \draw (1,0) circle (2pt);
      \filldraw [gray!50] (1.5,-.5) circle (2pt);
      \draw(1.5,-.5) circle (2pt);
      \end{tikzpicture}
      ~ + \hspace{1pt} \frac{1}{4} \hspace{2pt} ~
     \begin{tikzpicture}[baseline=-.1cm]
     \draw (3,0) circle (.5cm);
     \draw (2.5,0) to[out=30,in=150] (3.5,0);
     \filldraw [gray!50] (3,.5) circle (2pt);
     \draw(3,.5) circle (2pt);
     \draw (2.5,0) to[out=-30,in=210] (3.5,0);
     \filldraw [gray!50] (3.5,0) circle (2pt);
     \draw (3.5,0) circle (2pt);
     \filldraw [gray!50] (2.5,0) circle (2pt);
     \draw (2.5,0) circle (2pt);
     \filldraw [gray!50] (3,-.5) circle (2pt);
     \draw(3,-.5) circle (2pt);
     \end{tikzpicture}\\
%
     \gamma_{ij}\phi_i\phi_j = &
     + \frac{1}{4} ~
     \begin{tikzpicture}[baseline=-.1cm]
     \draw (0,0) circle (.5cm);
     \draw (-.5,0) to [out=0,in=180] (.5,0);
     \filldraw [gray!50] (-.5,0) circle (2pt);
     \draw (-.5,0) circle (2pt);
     \filldraw [gray!50] (.5,0) circle (2pt);
     \draw(.5,0) circle (2pt);
     \end{tikzpicture}
     \, \, - \frac{1}{16} ~ \,
     \begin{tikzpicture}[baseline=-.1cm]
     \draw (0,0) circle (.5cm);
     \draw (0,.5) to [out=-90,in=30] (-.469,-.171);
     \draw (0,.5) to [out=-90,in=150] (.469,-.171);
     \filldraw [gray!50] (0,.5) circle (2pt);
     \draw (0,.5) circle (2pt);
     \filldraw [gray!50] (.469,-.171) circle (2pt);
     \draw (.469,-.171) circle (2pt);
     \filldraw [gray!50] (-.469,-.171) circle (2pt);
     \draw (-.469,-.171) circle (2pt);
     \end{tikzpicture}
     \end{split}
\end{equation}
The system of beta functions of \eqref{eq:marginalV4} is
\begin{equation}\label{eq:betasystem-dc4}
  \begin{split}
  \beta_{\lambda_1} & = -\lambda_1 \epsilon +\frac{2}{3} \lambda_1 (4 \lambda_1+3
  \lambda_2) -\frac{1}{3} \lambda_1 \left(14 \lambda_1^2+22 \lambda_1
   \lambda_2+5 \lambda_2^2\right)\\
   & +\frac{1}{108} \lambda_1 \left[8 \lambda_1^3 (132 \zeta_3+185)+6 \lambda_1^2 \lambda_2 (384 \zeta_3+659)+27 \lambda_1 \lambda_2^2 (32
   \zeta_3+107)+756 \lambda_2^3\right] \,,\\
\beta_{\lambda_2} & = -\lambda_2 \epsilon +\lambda_2 (4 \lambda_1+3\lambda_2) -\frac{1}{9} \lambda_2 \left(82 \lambda_1^2+138 \lambda_1\lambda_2+51 \lambda_2^2\right)+\frac{1}{216} \lambda_2 \left[8 \lambda_1^3 (672\zeta_3+821) \right.\\
& +   \vphantom{\frac{a}{b}}  \left.54 \lambda_1^2 \lambda_2 (256 \zeta_3+325)+108 \lambda_1 \lambda_2^2
   (96 \zeta_3+131)+27 \lambda_2^3 (96 \zeta_3+145)\right]\,.
\end{split}
\end{equation}
Here and everywhere else in the paper, $\zeta_n$ refers to the Riemann zeta function.

\subsection{\texorpdfstring{\boldmath{$d_c=3$}}{beta dc=3}}

The RG flow of a theory in $d=3-\epsilon$
can be found in Ref.~\cite{Osborn:2017ucf} in fully functional form at NLO
(corresponding to four loops for the potential).
\begin{equation}
\begin{split}
        \beta_V  = &
        + \frac{1}{3} ~
        \begin{tikzpicture}[baseline=-.1cm]
        \draw (0,0) circle (.5cm);
        \draw (-.5,0) to [out=0,in=180] (.5,0);
        \filldraw [gray!50] (.5,0) circle (2pt);
        \draw (.5,0) circle (2pt);
        \filldraw [gray!50] (-.5,0) circle (2pt);
        \draw (-.5,0) circle (2pt);
        \end{tikzpicture}
        ~ +\frac{1}{6} ~
        \begin{tikzpicture}[baseline=-.1cm]
        \draw (1.5,0) circle (.5cm);
        \draw (1,0) to[out=50,in=130] (2,0);
        \draw (1,0) to[out=0,in=180] (2,0);
        \draw (1,0) to[out=-50,in=-130] (2,0);
        \filldraw [gray!50] (1.5,.5) circle (2pt);
        \draw(1.5,.5) circle (2pt);
        \filldraw [gray!50] (1,0) circle (2pt);
        \draw (1,0) circle (2pt);
        \filldraw [gray!50] (2,0) circle (2pt);
        \draw (2,0) circle (2pt);
        \end{tikzpicture}
        ~ - \frac{4}{3} ~
        \begin{tikzpicture}[baseline=-.1cm]
        \draw (3,0) circle (.5cm);
        \draw (2.531,-.171) to [out=-30,in=-150] (3.469,-.171);
        \draw (2.531,-.171) to [out=30,in=150] (3.469,-.171);
        \draw (3,.5) to[out=-90,in=150] (3.469,-.171);
        \filldraw [gray!50] (3,.5) circle (2pt);
        \draw (3,.5) circle (2pt);
        \filldraw [gray!50] (3.469,-.171) circle (2pt);
        \draw (3.469,-.171) circle (2pt);
        \filldraw [gray!50] (2.531,-.171) circle (2pt);
        \draw (2.531,-.171) circle (2pt);
        \end{tikzpicture}
        ~ - \frac{\pi}{12} ~
        \begin{tikzpicture}[baseline=-.1cm]
        \draw (4.5,0) circle (.5cm);
        \draw (4.031,-.171) to [out=-30,in=-150] (4.969,-.171);
        \draw (4.5,.5) to [out=-90,in=30] (4.031,-.171);
        \draw (4.5,.5) to[out=-90,in=150] (4.969,-.171);
        \filldraw [gray!50] (4.5,.5) circle (2pt);
        \draw (4.5,.5) circle (2pt);
        \filldraw [gray!50] (4.969,-.171) circle (2pt);
        \draw (4.969,-.171) circle (2pt);
        \filldraw [gray!50] (4.031,-.171) circle (2pt);
        \draw (4.031,-.171) circle (2pt);
     \end{tikzpicture} \\
%
      \gamma_{ij}\phi_i\phi_j = &
     + \frac{1}{90} ~
     \begin{tikzpicture}[baseline=-.1cm]
     \draw (1.5,0) circle (.5cm);
     \draw (1,0) to[out=50,in=130] (2,0);
     \draw (1,0) to[out=0,in=180] (2,0);
     \draw (1,0) to[out=-50,in=-130] (2,0);
     \filldraw [gray!50] (1,0) circle (2pt);
     \draw (1,0) circle (2pt);
     \filldraw [gray!50] (2,0) circle (2pt);
     \draw (2,0) circle (2pt);
     \end{tikzpicture}
     \end{split}
\end{equation}
The system of beta functions of \eqref{eq:marginalV3} is
\begin{equation}\label{eq:betasystem-dc3}
  \begin{split}
  \beta_{\upsilon_1} & = -2 \upsilon_1 \epsilon + \frac{4}{15} \left(22 \upsilon_1^2+12 \upsilon_1
   \upsilon_2+\upsilon_2^2\right)-\frac{1}{450} \left\{32 \left(826\!+\!85 \pi ^2\right)
   \upsilon_1^3+48 \upsilon_1^2 \left[21 \left(32\!+\!3 \pi ^2\right) \upsilon_2 \right.\right. \\
   & \vphantom{\frac{a}{b}} \left.+5 \left(35+3\pi ^2\right) \upsilon_3\right]+12 \upsilon_1 \left[\left(868+71 \pi ^2\right)
   \upsilon_2^2+30 \left(10+\pi ^2\right) \upsilon_2 \upsilon_3-40
   \upsilon_3^2\right]\\
   & \vphantom{\frac{a}{b}} \left.+\upsilon_2^2 \left[\left(960+74 \pi ^2\right) \upsilon_2+15
   \left(32+3 \pi ^2\right) \upsilon_3\right]\right\} \,,\\
   \beta_{\upsilon_2} & = -2\upsilon_2 \epsilon  +\frac{4}{15} \left[ 6 \upsilon_1 (6 \upsilon_2+5 \upsilon_3)+\upsilon_2 (13 \upsilon_2+10 \upsilon_3)\right] + \frac{1}{450} \left\{-24 \upsilon_1^2 \left[8 \left(301+32 \pi
   ^2\right) \upsilon_2 \right. \right. \\
   & \vphantom{\frac{a}{b}}\left. + 5 \left(532\!+\!51 \pi ^2\right) \upsilon_3\right] \!-\!3 \upsilon_1 \left[8
   \left(2424\!+\!241 \pi ^2\right) \upsilon_2^2+\!60 \left(476\!+\!41 \pi ^2\right) \upsilon_2
   \upsilon_3+225 \left(40\!+\!3 \pi ^2\right) \upsilon_3^2\right]\\
   & \vphantom{\frac{a}{b}} \left. -\upsilon_2 \left(4
   \left(3124+291 \pi ^2\right) \upsilon_2^2+15 \left(1376+117 \pi ^2\right) \upsilon_2
   \upsilon_3+15 \left(568+45 \pi ^2\right) \upsilon_3^2\right)\right\} \,, \\
   \beta_{\upsilon_3} & = -2 \upsilon_3 \epsilon +\frac{4}{45} \left[15 \upsilon_3 (4 \upsilon_1+5
   \upsilon_3)+32 \upsilon_2^2+120 \upsilon_2 \upsilon_3\right] + \frac{1}{450} \left\{-64 \left(371+45 \pi ^2\right) \upsilon_1^2
   \upsilon_3 \right. \\
   & \vphantom{\frac{a}{b}} -12 \upsilon_1 \left[256 \left(9+\pi ^2\right) \upsilon_2^2+4 \left(2284+255
   \pi ^2\right) \upsilon_2 \upsilon_3+5 \left(1244+135 \pi ^2\right) \upsilon_3^2\right] \\
   & \vphantom{\frac{a}{b}} -32
   \left(612+65 \pi ^2\right) \upsilon_2^3-8 \left(9934+1035 \pi ^2\right) \upsilon_2^2
   \upsilon_3-30 \left(3088+315 \pi ^2\right) \upsilon_2 \upsilon_3^2\\
   &\vphantom{\frac{a}{b}} \left.-15 \left(2248+225 \pi
   ^2\right) \upsilon_3^3\right\}\,.
  \end{split}
\end{equation}

\subsection{\texorpdfstring{\boldmath{$d_c=\frac{8}{3}$}}{beta dc=8/3}}

The functional form of the RG flow in $d=\frac{8}{3}-\epsilon$ can be obtained
by promoting the single-field case of Ref.~\cite{ODwyer:2007}
to arbitrary number of fields, which can be done unambiguously at LO and NLO
without further computations \cite{Zinati:2019gct}. In Sect.~\ref{sec:dc8/3}
we only use the LO, because inserting the full potential \eqref{eq:marginalV3} in
$\beta_V$ at NLO produces a rather demanding number of terms.
The single most expensive diagram produces
about $1.5\cdot 10^{7}$ by itself, that have to be simplified
iteratively through contractions of generalized deltas.
To explain the difficulty, consider that for decreasing $d_c$
both the number of contractions in the diagrams and the number of couplings increase,
making the study of each new multicritical RG exponentially more complicate.
\begin{equation}
\begin{split}
        \beta_V  = &
        + \frac{1}{8} ~
        \begin{tikzpicture}[baseline=-.1cm]
        \draw (0,0) circle (.5cm);
        \draw (-.5,0) to [out=50,in=130] (.5,0);
        \draw (-.5,0) to[out=-50,in=-130] (.5,0);
        \filldraw [gray!50] (-.5,0) circle (2pt);
        \draw (-.5,0) circle (2pt);
        \filldraw [gray!50] (.5,0) circle (2pt);
        \draw (.5,0) circle (2pt);
        \end{tikzpicture}
        ~ +\frac{1}{160} ~
        \begin{tikzpicture}[baseline=-.1cm]
        \draw (1.5,0) circle (.5cm);
        \draw (1,0) to [out=50,in=130] (2,0);
        \draw (1,0) to [out=75,in=180] (1.5,.375);
        \draw (1.5,.375) to [out=0,in=115] (2,0);
        \draw (1,0) to [out=-75,in=180] (1.5,-.375);
        \draw (1.5,-.375) to [out=0,in=-115] (2,0);
        \draw (1,0) to [out=0,in=180] (2,0);
        \draw (1,0) to[out=-50,in=-130] (2,0);
        \filldraw [gray!50] (1.5,.5) circle (2pt);
        \draw (1.5,.5) circle (2pt);
        \filldraw [gray!50] (1,0) circle (2pt);
        \draw (1,0) circle (2pt);
        \filldraw [gray!50] (2,0) circle (2pt);
        \draw (2,0) circle (2pt);
        \end{tikzpicture}
        ~ + \frac{9}{80} ~
        \begin{tikzpicture}[baseline=-.1cm]
        \draw (3,0) circle (.5cm);
        \draw (2.531,-.171) to [out=-30,in=-150] (3.469,-.171);
        \draw (2.531,-.171) to [out=30,in=150] (3.469,-.171);
        \draw (3,.5) to[out=-45,in=115] (3.469,-.171);
        \draw (2.531,-.171) to[out=65,in=180] (3,.15);
        \draw (3,.15) to[out=0,in=115] (3.469,-.171);
        \draw (2.531,-.171) to [out=0,in=180] (3.469,-.171);
        \filldraw [gray!50] (3,.5) circle (2pt);
        \draw (3,.5) circle (2pt);
        \filldraw [gray!50] (3.469,-.171) circle (2pt);
        \draw (3.469,-.171) circle (2pt);
        \filldraw [gray!50] (2.531,-.171) circle (2pt);
        \draw (2.531,-.171) circle (2pt);
        \end{tikzpicture}
        ~ - \frac{3}{8} ~
        \begin{tikzpicture}[baseline=-.1cm]
        \draw (4.5,0) circle (.5cm);
        \draw (4.031,-.171) to [out=-30,in=-150] (4.969,-.171);
        \draw (4.031,-.171) to [out=30,in=150] (4.969,-.171);
        \draw (4.5,.5) to[out=-40,in=110] (4.969,-.171);
        \draw (4.969,-.171) to[out=150,in=-90] (4.5,.5);
        \draw (4.031,-.171) to [out=0,in=180] (4.969,-.171);
        \filldraw [gray!50] (4.5,.5) circle (2pt);
        \draw (4.5,.5) circle (2pt);
        \filldraw [gray!50] (4.969,-.171) circle (2pt);
        \draw (4.969,-.171) circle (2pt);
        \filldraw [gray!50] (4.031,-.171) circle (2pt);
        \draw (4.031,-.171) circle (2pt);
        \end{tikzpicture}\\
        & -\frac{\Gamma\left(\frac{1}{3}\right)^3}{24} ~
        \begin{tikzpicture}[baseline=-.1cm]
       \draw (6,0) circle (.5cm);
       \draw (5.531,-.171) to [out=-30,in=-150] (6.469,-.171);
       \draw (6,.5) to[out=-40,in=110] (6.469,-.171);
       \draw (6,.5) to[out=-90,in=30] (5.531,-.171);
       \draw (6,.5) to[out=220,in=70] (5.531,-.171);
       \draw (6.469,-.171) to[out=150,in=-90] (6,.5);
       \filldraw [gray!50] (6,.5) circle (2pt);
       \draw (6,.5) circle (2pt);
       \filldraw [gray!50] (6.469,-.171) circle (2pt);
       \draw (6.469,-.171) circle (2pt);
       \filldraw [gray!50] (5.531,-.171) circle (2pt);
       \draw (5.531,-.171) circle (2pt);
       \end{tikzpicture}
       ~ + \frac{3}{64} \left[\sqrt{3} \pi -3
          (2+\log 3)\right] ~
       \begin{tikzpicture}[baseline=-.1cm]
       \draw (7.5,0) circle (.5cm);
       \draw (7.031,-.171) to [out=-30,in=-150] (7.969,-.171);
       \draw (7.031,-.171) to [out=30,in=150] (7.969,-.171);
       \draw (7.5,.5) to[out=-90,in=30] (7.031,-.171);
       \draw (7.969,-.171) to[out=150,in=-90] (7.5,.5);
       \draw (7.031,-.171) to [out=0,in=180] (7.969,-.171);
       \filldraw [gray!50] (7.5,.5) circle (2pt);
       \draw (7.5,.5) circle (2pt);
       \filldraw [gray!50] (7.969,-.171) circle (2pt);
       \draw (7.969,-.171) circle (2pt);
       \filldraw [gray!50] (7.031,-.171) circle (2pt);
       \draw (7.031,-.171) circle (2pt);
       \end{tikzpicture}\\
%
      \gamma_{ij}\phi_i\phi_j = &
       + ~ \frac{1}{2240} ~~
       \begin{tikzpicture}[baseline=-.1cm]
       \draw (1.5,0) circle (.5cm);
       \draw (1,0) to [out=50,in=130] (2,0);
       \draw (1,0) to [out=75,in=180] (1.5,.375);
       \draw (1.5,.375) to [out=0,in=115] (2,0);
       \draw (1,0) to [out=-75,in=180] (1.5,-.375);
       \draw (1.5,-.375) to [out=0,in=-115] (2,0);
       \draw (1,0) to [out=0,in=180] (2,0);
       \draw (1,0) to[out=-50,in=-130] (2,0);
       \filldraw [gray!50] (1,0) circle (2pt);
       \draw (1,0) circle (2pt);
       \filldraw [gray!50] (2,0) circle (2pt);
       \draw (2,0) circle (2pt);
       \end{tikzpicture}
     \end{split}
\end{equation}
The LO beta functions of \eqref{eq:marginalV8/3} are
\begin{equation}\label{eq:betasystem-dc8/3}
  \begin{split}
  \beta_{\rho_1} & = -3 \rho_1
  \epsilon + \frac{1}{35} \left[ 2144 \rho_1^2+\rho_1 (912 \rho_2+90 \rho_3+72 \rho_4)+3 \rho_2 (24 \rho_2+5 \rho_3)\right] \,, \\
  \beta_{\rho_2} & = -3 \rho_2\epsilon +\frac{1}{70}\! \left[8 \rho_1 (920 \rho_2\!+\!525 \rho_3\!+\!444
   \rho_4\!+\!105 \rho_5)\!+1936 \rho_2^2 + 4 \rho_2 (340 \rho_3\!+\!204
   \rho_4\!+\!35 \rho_5) \right. \\
   & \vphantom{\frac{a}{b}} \left. +15 \rho_3 (15 \rho_3+4 \rho_4)\right] \,, \\
   \beta_{\rho_3} & = -3 \rho_3 \epsilon + \frac{1}{105} \left[480 \rho_1 (16 \rho_3+12 \rho_4+21
   \rho_5)+3200 \rho_2^2+24 \rho_2 (335 \rho_3+244 \rho_4+245
   \rho_5)\right.\\
   & \vphantom{\frac{a}{b}} \left.\vphantom{\frac{a}{b}}+3075 \rho_3^2+3600 \rho_3 \rho_4+3150 \rho_3 \rho_5+384
   \rho_4^2\right] \,, \\
   \beta_{\rho_4} & = -3 \rho_4 \epsilon + \frac{1}{420} \left(13440 \rho_1 \rho_4+3200 \rho_2^2+3000
   \rho_2 \rho_3+11232 \rho_2 \rho_4+840 \rho_2 \rho_5+225
   \rho_3^2 \right.\\
   &\left. \vphantom{\frac{a}{b}} + 5160 \rho_3 \rho_4+4224 \rho_4^2+1680 \rho_4\rho_5\right) \,, \\
   \beta_{\rho_5} & = -3 \rho_5 \epsilon + \frac{1}{35} \left(1120 \rho_1 \rho_5+1200 \rho_2
   \rho_3+1600 \rho_2 \rho_4+2800 \rho_2 \rho_5+1275 \rho_3^2+3240
   \rho_3 \rho_4 \right.\\
   & \vphantom{\frac{a}{b}} \left. +3850 \rho_3 \rho_5+1968 \rho_4^2+4760 \rho_4
   \rho_5+2450 \rho_5^2\right)\,.
  \end{split}
\end{equation}

\subsection{\texorpdfstring{\boldmath{$d_c=\frac{10}{3}$}}{beta dc=10/3}}

The functional flow of a multiscalar theory in $d=\frac{10}{3}-\epsilon$
can be obtained by promoting the single field case found in \cite{Codello:2017}
to an arbitrary number of flavors. This flow is presently known only to the leading order.
\begin{equation}
\begin{split}
        \beta_V  = &
        + \, \frac{3}{4} ~ \, \,
        \begin{tikzpicture}[baseline=-.1cm]
        \draw (0,0) circle (.5cm);
        \draw (-.5,0) to [out=30,in=150] (.5,0);
        \draw (-.5,0) to[out=-30,in=-150] (.5,0);
        \filldraw [gray!50] (0,.5) circle (2pt);
        \draw (0,.5) circle (2pt);
        \filldraw [gray!50] (.5,0) circle (2pt);
        \draw (.5,0) circle (2pt);
        \filldraw [gray!50] (-.5,0) circle (2pt);
        \draw (-.5,0) circle (2pt);
        \end{tikzpicture}
        ~ - \frac{27}{8} ~
        \begin{tikzpicture}[baseline=-.1cm]
        \draw (1.5,0) circle (.5cm);
        \draw (1.5,.5) to[out=-90,in=30] (1.067,-.25);
        \draw (1.5,.5) to[out=-90,in=150] (1.933,-.25);
        \filldraw [gray!50] (1.5,.5) circle (2pt);
        \draw(1.5,.5) circle (2pt);
        \filldraw [gray!50] (1.933,-.25) circle (2pt);
        \draw (1.933,-.25) circle (2pt);
        \filldraw [gray!50] (1.067,-.25) circle (2pt);
        \draw (1.067,-.25) circle (2pt);
        \end{tikzpicture}\\
%
       \gamma_{ij}\phi_i\phi_j = &
        + \frac{3}{80} ~
        \begin{tikzpicture}[baseline=-.1cm]
        \draw (0,0) circle (.5cm);
        \draw (-.5,0) to [out=30,in=150] (.5,0);
        \draw (-.5,0) to[out=-30,in=-150] (.5,0);
        \filldraw [gray!50] (.5,0) circle (2pt);
        \draw (.5,0) circle (2pt);
        \filldraw [gray!50] (-.5,0) circle (2pt);
        \draw (-.5,0) circle (2pt);
        \end{tikzpicture}
     \end{split}
\end{equation}
The beta functions of \eqref{eq:marginalV103} are
\begin{equation}\label{eq:betasystem-dc10/3}
  \begin{split}
    \beta_{\kappa_1} & = -\frac{3}{2} \kappa_1 \epsilon - \frac{229}{6} \kappa_1^3 \\
    \beta_{\kappa_2} & = -\frac{3}{2} \kappa_2 \epsilon  - \frac{1}{2250}
    \left\{15525 \kappa_1^2 \kappa_2 + 50 \kappa_1 \left[ 216
   \kappa_2^2+27 \kappa_2 (2 \kappa_3+3 \kappa_4) + 4 \left(2 \kappa_3^2+6
   \kappa_3 \kappa_4+3 \kappa_4^2\right)\right] \right.\\
   & \vphantom{\frac{a}{b}} + 2139 \kappa_2^3 + 1350 \kappa_2^2
   (2 \kappa_3+3 \kappa_4) + 2 \kappa_2 \left(1072 \kappa_3^2+3216 \kappa_3
   \kappa_4+1743 \kappa_4^2\right) \\
   & \vphantom{\frac{a}{b}}  \left. + 40 \left(44 \kappa_3^3+198 \kappa_3^2
   \kappa_4+219 \kappa_3 \kappa_4^2+72 \kappa_4^3\right)\right\} \\
   \beta_{\kappa_3} & = -\frac{3}{2}\kappa_3 \epsilon + \frac{1}{4500}\left\{150 \kappa_1^2 \kappa_3-75 \kappa_1 \kappa_2 (27
   \kappa_2-16 \kappa_3)-2 \left[675 \kappa_2^3+3 \kappa_2^2 (536 \kappa_3+135
   \kappa_4) \right.\right.\\
   & \vphantom{\frac{a}{b}} \left.\left.  +180 \kappa_2 \left(22 \kappa_3^2+14 \kappa_3
   \kappa_4-\kappa_4^2\right)+2 \kappa_3 \left(2404 \kappa_3^2+3852 \kappa_3
   \kappa_4+1161 \kappa_4^2\right)\right]\right\} \\
   \beta_{\kappa_4} & = -\frac{3}{2}  \kappa_4\epsilon  -\frac{1}{750} \kappa_4 \left(-25 \kappa_1^2-200 \kappa_1
   \kappa_2+446 \kappa_2^2+2080 \kappa_2 \kappa_3+1500 \kappa_2 \kappa_4+3096
   \kappa_3^2 \right. \\
   & \left. \vphantom{\frac{a}{b}}  + 5148 \kappa_3 \kappa_4+1974 \kappa_4^2\right)
  \end{split}
\end{equation}

\section{\texorpdfstring{Critical theory in $d=4-\epsilon$}{Critical theory in 4-epsilon}}
\label{sec:dc4}

The critical behavior of a quenched ferromagnet in $d=4-\epsilon$ dimensions comes from truncating  Eq.~\eqref{eq:cumulantsexp} to the second cumulant, which gives the following marginal potential
\begin{equation}\label{eq:marginalV4}
\begin{split}
  V(\phi_a) &= g_1 ~ \sum_a E_a^2(x) + g_2 ~ \sum_{a \neq b} E_a(x) E_b(x)
= \lambda_1 ~\Bigl(\sum_a \phi_a^2\Bigr)^2 + \lambda_2 ~ \sum_a \phi_a^4 \,.
\end{split}
\end{equation}
The second form of the potential might be more familiar to some readers,
because it makes explicit its field-theoretical nature.
The couplings in the two parametrizations are easily related,
$g_1=\lambda_1+\lambda_2$ and $g_2=2\lambda_1$,
with the added value that one immediately recognizes $O(N)$ as
an the enhanced (maximal) symmetry group that is achieved if $\lambda_2=0$.
The RG flow that we need for the potential of this model is given in \eqref{eq:betasystem-dc4}.
For $N>0$, it admits two interesting fixed points:
the $O(N)$-invariant model, and the $H_N$-invariant one.
In the limit $N\to 0$, the $O(N)$-invariant fixed point has been associated to the universality class
of the self-avoiding walk (SAW) and we suggest \cite{Wiese:2020} for a recent discussion.

The limit $N\to 0$ of the $H_N$-invariant point is more tricky, because
critical exponents and fixed point coordinates are divergent for $N=0$,
and therefore cannot be trivially continued.
The reason why this happens is that an expansion in powers of $\epsilon^{1/2}$,
rather than $\epsilon$ itself, becomes appropriate due to a degeneracy in the leading solution of the two beta functions.
Adopting the correct expansion, one finds
the so-called random Ising FP (a.k.a.\ Khmelnitskii FP) of order $\epsilon^{1/2}$ \cite{Grinstein:1976,khmel1975}, with coordinates at three-loop order given by
\begin{equation}\label{eq:4dfp}
\begin{split}
  \lambda_1^{\star} & = -3 \sqrt{\frac{3}{106}} ~\epsilon^{1/2}~ +\frac{9 (63 \zeta_3 + 110)}{2809} ~\epsilon \,,
  \\
  \lambda_2^{\star} & = +2 \sqrt{\frac{6}{53}} ~\epsilon^{1/2}~
                  -\frac{36 (21 \zeta_3 + 19)}{2809} ~\epsilon~
                  -\frac{18 \sqrt{\frac{6}{53}} (31 \zeta_3 + 23)}{2809} ~\epsilon ^{3/2} \,.
\end{split}
\end{equation}
This fixed point is the first example of new \emph{random} universality class,
which emerges in the presence of (weak) quenched disorder. There is actually an
additional fixed point that solves the same system in powers of $\epsilon^{1/2}$,
but it is unstable.

For completeness, we report here the corresponding three-loop critical exponents $\eta$ and $\nu$ (they can be found in Ref.~\cite{Jayaprakash:1977}), which read
\begin{equation}
\begin{split}
  \eta  &= - \frac{1}{106}~\epsilon~ + \frac{9 \sqrt{\frac{6}{53}} (7 \zeta_3+24)}{2809}~\epsilon ^{3/2} \,, \\
  \nu & = \frac{1}{2} + \frac{1}{2} \sqrt{\frac{3}{106}} ~\epsilon^{1/2}~ - \frac{(756 \zeta_3 - 535)}{22472}~ \epsilon  \,,
\end{split}
\end{equation}
while the scaling dimensions associated to the energy operator, to the quadratic deformation $Y$ and the scaling dimension $\Delta'_E$ characterizing the logarithmic correction are given respectively by
\begin{eqnarray}
  \Delta_E  &=& 2 + \sqrt{\frac{6}{53}} ~\epsilon^{1/2} - \frac{(756 \zeta_3 + 5401)}{5618}  ~\epsilon \,,
  \qquad
   \Delta_Y = 2 - \sqrt{\frac{6}{53}} ~\epsilon^{1/2} + \frac{(756 \zeta_3 - 4033)}{5618}  ~\epsilon \,,
 \nonumber \\
  \Delta'_E & =& -\sqrt{\frac{3}{106}} ~\epsilon^{1/2} + \frac{9 (21 \zeta_3 + 19)}{2809} ~\epsilon  \,.
\end{eqnarray}
An interesting aspect of the random Ising critical point is that for $d=3$ the anomalous dimension is positive
according to lattice simulations, which implies that the perturbative result should change sign upon resummation
to $\epsilon=1$. This is of course very nontrivial and deserving of more investigations with nonperturbative tools.

We include the values for the functions $a$ and $c_T$ at the random fixed point
as are defined in Sect.~\ref{sec:method}
\begin{equation}
\begin{split}
  a^{\star} = \frac{3}{212}~\epsilon ^2 \,,\qquad\qquad
  c_T ^{\star}  = 1 +\frac{5}{636}~\epsilon -\frac{15 \sqrt{\frac{3}{106}} (7 \zeta_3 + 24)}{2809}~\epsilon ^{3/2}\,.
\end{split}
\end{equation}
We finally give the quantities related to the fractal dimension and the crossover exponents,
\begin{equation}
\begin{split}
  d_f & = 2 -\sqrt{\frac{6}{53}} ~\epsilon^{1/2} + \frac{7 (108 \zeta_3 - 31)}{5618}  ~\epsilon \,,
 \\
  \phi_Y & = 1+\sqrt{\frac{6}{53}} ~\epsilon^{1/2} -\frac{3(126 \zeta_3 + 61)}{2809} ~\epsilon -\frac{\sqrt{\frac{3}{106}} (1512 \zeta_3 + 2269)}{5618} ~\epsilon ^{3/2}\,.
\end{split}
\end{equation}

The interpretation of the fixed point \eqref{eq:4dfp} as being the one pertaining to the random system
can be strengthened by a couple of observations.
From the cumulant expansion, we expect the coupling $g_2$ to be negative, since it is proportional to the negative of the variance of the probability distribution function. This expectation is corroborated by the analysis proposed in Refs.~\cite{Lubensky1975, Aharony1976}, in which it is confirmed that $g_2$ cannot change sign under renormalization, providing a ``physical'' region in parameters' space where to expect the non-trivial random universality class.
Moreover, since we are interested in phase transitions that can be second order in the pure systems,
a requirement on the fixed point could be that the coupling corresponding to the original pure system is positive,
which is confirmed by the value of $g_1$.
The above discussion allows us to disregard an unphysical fixed point, in fact,
the system of beta functions of Eq.~\eqref{eq:marginalV4} admits another fixed point which however lies in the ``unphysical'' region $\{ g_1 < 0, g_2 > 0 \}$.
A similar logic is applied in Sects.~\ref{sec:dc3} to \ref{sec:dc8/3} and proves very useful
when deciding which fixed point is physically meaningful.


\bibliographystyle{jhep.bst}
\bibliography{RDI}

\end{document}